# The Spectral Coupled Cluster Method with $k$ dependency for strongly correlated lattices.

Alessandro Mirone
European Synchrotron Radiation Facility, BP 220, F-38043 Grenoble Cedex, France

We adapt the Coupled Cluster Method to solid state strongly correlated lattice Hamiltonians extending the Coupled Cluster linear response method to the calculation of electronic spectra and obtaining the space-time Fourier transforms of generic Green's functions. We apply our method to the $MnO_2$ plane with orbital and magnetic ordering, to interpret electron energy loss experimental data, and to the Hubbard model, where we get insight into the pairing mechanism.

## 1. INTRODUCTION

The recent developments of the Linear Response Coupled Cluster[1] have stretched the applicability of the Coupled Cluster Method[2] (CCM), originally formulated as a ground state approximation, to excited states. In particular Crawford and Ruud have calculated vibrational eigenstates contributions to Raman optical activity[3] while Govind et al. have calculated excitonic states in potassium bromide[4]. More recently the Spectral Coupled Cluster Method (SCCM) has been derived from the linear response to calculate electronic spectra. This is done by iteratively refining a resolvent operator which describes the response of the Coupled Cluster solution to probe operators and then by obtaining the spectral functions by the diagrammatical expansion of their expectation value[5].

In this work we further expand SCCM for solid state model Hamiltonians, by adapting it to the calculation of space-time Fourier transforms of generic Green's functions. In the next section (2) we recall first the CC formalism and then introduce the new formulation of SCCM with $k$ dependency.

With the goal of simulating the energy loss spectra of the orbitally ordered $MnO_2$ plane in $La_{0.5}Sr_{1.5}MnO_4$ we detail, in section (3), the model Hamiltonian that we use to simulate the $MnO_2$ plane, and also two simplified forms of this Hamiltonian which describe the Hubbard model and a simple linear chain that we can solve exactly for a preliminary validation of our SCCM implementation. We discuss in section (4) the results for the simple linear chain, for the Hubbard model and, finally,for the $MnO_2$ plane. For the Hubbard model we compare our calculation to the exact resolution of the $4\times 4$ Hubbard system, getting insight into the role of the collective spinonic excitations in the ground state and a possible pairing mechanism. For the orbitally ordered $MnO_2$ plane we simulate, by calculating the dynamic structure factor, the low energy resonances observed in an electron energy loss experiment.

## 2. METHOD

The Coupled Cluster method[2] uses an exponential ansatz to preserve extensive properties. Given a reference state $|\Phi_0\rangle$, the solution $|\Psi\rangle$ is represented, in this ansatz, by :

$$|\Psi\rangle = e^{S}|\Phi_0\rangle \simeq e^{S_N}|\Phi_0\rangle \tag{1}$$

where $S$ gives the ideal exact solution and $S_N$ is a sum, truncated to $N$ terms, of products of electron-hole pair excitations:

$$S_N = \sum_{i=1}^{N} t_i \; Symm\left\{\prod_{k=1}^{n_i} \hat{c}^{\dagger}_{\alpha_{i,k}}\hat{c}^{\dagger}_{a_{i,k}}\right\} \tag{2}$$

Each term in the sum is the product of a set of electron(hole)-creation operators $\hat{c}^{\dagger}$, and is determined by a choice of indexes $\alpha_{i,k}(\, a_{i,k})$, with the greek( latin) letter $\alpha(\, a)$ ranging over empty( occupied) orbitals of the reference state$|\Phi_0\rangle$. One can consider the reference state as the *vacuum* state and that each term $i$ in the sum $S_N$ creates, from the vacuum, an excited state which is populated by $n_i$ holes and $n_i$ electrons. The *Symm* operator makes the ansatz symmetric for the Hamiltonian symmetry subgroup which transforms, up to a factor, the reference state $|\Phi_0\rangle$ into itself. The $t_i$'s are free coefficients that are obtained from the Schrodinger's eigen-equation:

$$H|\Psi\rangle - E|\Psi\rangle = 0 \tag{3}$$

by setting the eigen-equation residue to zero in the space of excited states which enter the $S_N$ sum:

$$0 = \langle\Phi_0|\left(\prod_{k=1}^{n_i} \hat{c}_{a_{i,k}}\hat{c}_{\alpha_{i,k}}\right) e^{-S_N}He^{S_N}|\Phi_0\rangle \;\; \forall i \in [1,N] \tag{4}$$

This equation enforces the orthogonality between $e^{-S_N}He^{S_N}|\Phi_0\rangle$ and the basis of excited states. The overlap with the reference state is non-zero and gives an approximation of the energy:

$$E \simeq E_N = \langle\Phi_0|e^{-S_N}He^{S_N}|\Phi_0\rangle \tag{5}$$

$$\tag{6}$$

The Coupled Cluster Method expands these equations by means of the Baker-Campell-Hausdorff expansion formula[6] which for two arbitrary operators $A$ and $B$ states that:

$$e^{-A}Be^{A} = B + [B,A] + 1/2[[B,A],A] + ..1/n![[...[B,A]...],A] + ... \tag{7}$$

The numerical applicability of CCM relies on the fact that, when $A$ is substituted with $S_N$ which contains creation operators only, and B is replaced by the Hamiltonian $H$ with the properties described below, then only the first five terms in the series can be non zero. This can be demonstrated considering that each term of $H$ is formed by a limited number of annihilation operators. The Coulomb interaction, in fact, contains products of four one-particle operators.

The accuracy of the CCM solution increases with $N$. To increase $N$, at each iteration, the excitation not yet contributing to $S_N$ and giving the largest scalar product in equation (4) is added, for the following step, to the list of excitations concurring to $S_{N+1}$. Once we have obtained the CCM ground state and its ground energy $E$, we are interested in the Green's functions that are obtained applying all sorts of probe operators on the ground state.

We are interested here in solid states model Hamiltonians applied on systems having periodic boundary conditions, and being composed of a Bravais lattice of identical unit cells. Each unit cell is composed of one or more atoms. The basis states of the Fock space are formed as antisymmetrised tensor products of atomic localised orbitals. Considering Hamiltonians that are invariant under translation by a Bravais lattice vector, and which are also time independent, it is convenient, for calculation purposes, to study the Green's functions space-time Fourier transform. All kinds of Green's functions (advanced, retarded, time-ordered, correlation functions) can be obtained from the following function (taking $\hbar = 1$ for notational simplification):

$$G(B,A,\omega,q,\gamma) = \frac{\langle\Phi_0|\, e^{S^\dagger} (P_q(B))^\dagger (H-\omega-i\gamma)^{-1} P_q(A) e^S \,|\Phi_0\rangle}{\langle\Phi_0|\, e^{S^\dagger} e^S \,|\Phi_0\rangle} \tag{8}$$

In this expression A and B represent, each, a product or a sum of products of $\hat{c}^\dagger$ and $\hat{c}$ operators, and $P_q$ is an operator which symmetrizes its argument under the Bravais lattice vector translations operators , here noted as $T_R$:

$$P_k(D) = \sum_R exp(ikR) T_R(D). \tag{9}$$

As an example of the different choices of $A$ and $B$, we can consider the adaptation of equation 8 to the dynamic structure factor calculation. The dynamic structure factor is observed experimentally in different kind of spectroscopies like electron energy loss (EELS) and Inelastic X-Ray Scattering Spectroscopy (IXSS). The dynamic structure factor is given by the space and time Fourier transform of the density-density correlation function[7,8] which is defined, writing $n(q,t)$ for the spatial Fourier component $q$ of the electron density at time $t$, as:

$$S(q,\omega) = \frac{1}{2\omega} \int_{-\infty}^{\infty} dt exp(i\omega t) \left\langle n^\dagger(q,t) n(q,0) \right\rangle \tag{10}$$

and corresponds, in equation 8, to the spectral density of the Green's function :

$$S(-q,\omega) = \frac{1}{2i\omega} (G(N_q,N_q,\omega,q,\gamma) - G(N_q,N_q,\omega,q,-\gamma)) \tag{11}$$

where $A$ and $B$ have been replaced by $N_q$, written summing over all the orbitals $i$ belonging to an unit cell $uc$ (at variance with $n_q$ which is summed over the whole lattice), and located at position $r_i$ inside $uc$:

$$N_q = \sum_{i \in uc} exp(iqr_i) c_i^\dagger c_i \tag{12}$$

where $c_i^\dagger$ denotes the operator which creates an electron on orbital $i$, not to be confused with $\hat{c}^\dagger$ which is defined with respect to the reference state taken as vacuum. In this work we will consider also, numerically, the partial contributions $S^{\uparrow\uparrow}, S^{\uparrow\downarrow}, S^{\downarrow\uparrow}, S^{\downarrow\downarrow}$ to $S(q,\omega)$ obtained by splitting $N_q = N_q^\uparrow + N_q^\downarrow$. Another example is the spectral function $A(q,\omega)$, which is the imaginary part of the one-particle retarded Green's function, related to angle resolved photo-emission spectroscopy (ARPES) and to inverse photo-emission spectroscopy (IPES):

$$A(q,\omega) = \frac{1}{2i\omega} (G(C_q,C_q,\omega,q,\gamma) - G(C_q,C_q,\omega,q,-\gamma)) \tag{13}$$

where $C_q$ is given, for $\omega > \mu$ (with $\mu$ the chemical potential, where we take the ground energy as the zero energy), by

$$C_q^{arpes} = \sum_{i \in uc} exp(iqr_i) c_i^\dagger \tag{14}$$

and, for $\omega < \mu$, by

$$C_q^{ipes} = \sum_{i \in uc} exp(-iqr_i) c_i \tag{15}$$

In the general case, to calculate equation (8), we have to solve two problems. The first one is to find an approximated solution $R$ for the resolvent equation :

$$(H-\omega-i\gamma)\,|R\rangle = P_q(A) e^S \,|\Phi_0\rangle \tag{16}$$

and the second one is to calculate the scalar product. The operator $P_q(A)$ can contain, for a generic $A$, both annihilation and creation operators. In order to better accommodate equation (16) into the framework of the coupled cluster method, which deals with excitations which are written as a product of creation operators only, we rewrite $A$ in a commutated form $A_S$ where annihilation operators disappear:

$$(H-\omega-i\gamma)\,|R\rangle = e^S P_q(A_S)\,|\Phi_0\rangle \tag{17}$$

where $A_S$ is defined as the Hausdorff expansion (7) of $e^{-S} A e^S$ discarded of all those terms having, with respect to the reference vacuum $|\Phi_0\rangle$, non-contracted annihilation operators:

$$e^S P_q(A_S)\,|\Phi_0\rangle = e^S P_q(e^{-S} A e^S)\,|\Phi_0\rangle\,. \tag{18}$$

The same transformation can be applied to the $B$ operators which appear in the Green's function equation (8). Note that in the above expression we have made use of the translational invariance of the $S$ operator.

We represent an approximated solution for the resolvent, introducing the approximating operator $R_{\omega,\gamma}$ and the following ansatz, which is refined at each increment of the number $N^r$ by adding a new excitation of the form $\prod_{k=1}^{n} \hat{c}_{j_k}^\dagger$:

$$|R\rangle = R_{\omega,\gamma} e^S \,|\Phi_0\rangle \tag{19}$$

$$R_{\omega,\gamma} = r_{\omega,\gamma}^0 P_q(A_S) + \textstyle\sum_{i=1}^{N^r} r_{\omega,\gamma}^i \left( P_q \left( \prod_{k=1}^{n_i^r} \hat{c}_{j_{i,k}}^\dagger \right) \right)_{\perp_i} \tag{20}$$

In this expression $r^i$ are free parameters and the $(.)_{\perp_i}$ symbol stands for the orthogonalisation of its argument with respect to $P_q(A_S)$ and to all the operators $P_q\left(\prod_{k=1}^{n_l^r}\hat{c}^\dagger_{j_{l,k}}\right)$ with $l < i$. We build our spectral CCM equations (SCCM equations) by multiplying equation (17) at the left with $e^{-S}$, and by setting the residue to zero :

$$0 \underset{\forall i\in[1,N^r]}{=} \langle\Phi_0| \left( P_q\left(\prod_{k=1}^{n_i^r}\hat{c}^\dagger_{j_{i,k}}\right)_{\perp_i}\right)^\dagger e^{-S}(H-\omega-i\gamma)R_{\omega,\gamma}e^S|\Phi_0\rangle \tag{21}$$

The equation (21) is expanded by the Hausdorff (7) expansion formula substituting $A$ with $S_N$ and $B$ with $(H-\omega-i\gamma)R_{\omega,\gamma}$. The Hausdorff expansion contains a finite number of non-zero terms because $R_{\omega,\gamma}$ contains only creation operators while each term of $H$ contains a maximum of four annihilation operators.

The expansion gives a set of linear equations for the $r$ parameters. The above equation (21) is the equivalent of equation 15 of Crawford and Ruud[3] transposed to the case of a periodic crystal and for excitations which diagonalise the translational symmetry operators.

The resolvent equation accuracy is improved by systematically increasing $N^r$, selecting, at each iteration, the set of numbers

$$\left\{\left(n^r_{N^r+1}, j_{N^r+1,0}, \ldots, j_{N^r+1,n^r_{N^r+1}}\right)\right\} \tag{22}$$

not yet contributing to the resolvent and corresponding to the largest scalar product in the SCCM equation (21). When we calculate the residue we fix $\omega=\omega_r$ at the center of the spectral region of interest. The parameters $r$ are given, over the spectral region of interest, by a linear algebra operations of the kind $r=(M_1)^{-1}(M_2+\omega M_3)$ where the $M$'s are matrices obtained from SCCM expansion. Once we know the $R$ operator we can calculate the spectra with the following equation:

$$G(A,B,q,\omega,\gamma) = \frac{\langle\Phi_0|\, e^{S^\dagger}\,(P_q(B_S))^\dagger R_{\omega,\gamma}e^S|\Phi_0\rangle}{\langle\Phi_0|\, e^{S^\dagger}e^S|\Phi_0\rangle} \tag{23}$$

This expression is calculated following the procedure already described in SCCM[5], using the Wick's theorem and the linked-cluster theorem to obtain a sum of products of Green's function of different orders, and a hierarchical set of Dyson equations for the Greens functions, that we truncate at the Hartree-Fock approximation.

## 3. MODEL

In this paper we extend our original application[5] of SCCM, which was tested on a small $2x2$ $MnO_2$ lattice with periodic boundary conditions, to a larger lattice. Considering a larger lattice allows us to finely resolve the $k$ dependency of the excitations and to accommodate in the model different choices of charge, spin, and orbital ordering. We detail below the model Hamiltonian used for the $MnO_2$ plane. We also validate our implementation, deriving from it simpler systems that can be solved exactly: the Hubbard $4\times4$ model and a simple $MnO$ chain. We perform these cross validations by keeping the same implementation, the same model, and by simply adapting the system parameters to the test cases.

### 3.1. Model and parameters for the $MnO_2$ plane

We keep in this work the same model Hamiltonian as in the previous paper[9] apart for its size and for a modification to the spin dependent term at the $Mn$ sites which is aimed to implement the spin ordering. For sake of clarity we report here all the details of the model Hamiltonian.

The $Mn$ sites are placed at integer coordinates $(2i,2j)$ with $j\in[0,N_x-1]$ and $i\in[0,N_y-1]$. The oxygen atoms are at positions $(2i+1,2j)$ and $(2i,2j+1)$. Periodic boundary conditions are applied, with periods $2N_y, 2N_x$ in the two directions. In order to limit the computational cost, we restrict the degrees of freedom to those orbitals which are the most important for the electronic properties of the $MnO_2$ plane in the manganites. These are the $e_g$ $3d$ orbitals of $Mn$, namely the $x^2-y^2$ and $3z^2-r^2$ orbitals, and the $p$ oxygen orbitals which point toward $Mn$ sites. The $x,y$ axis are in the $MnO_2$ plane and $z$ is out of plane. The $Mn$ $t_{2g}$ orbitals $(xy,xz,yz)$ of a given spin direction (up or down depending on the atomic site) have, each, occupancy fixed to 1. The other $t_{2g}$ of opposite spin direction have their occupancy freezed to 0.

We freeze the $t_{2g}$ degrees of freedom with the aim of reducing the computational cost of the model. The justification for this freezing is that, neglecting transition to (much) higher energy orbitals, the electron-electron interaction acting between two $t_{2g}$ electrons remains within the $t_{2g}$ sector, the one between two $e_g$'s: within the $e_g$ one. This is true for symmetry reasons. For the same reasons the interaction between a $t_{2g}$ electron and a $e_g$ one does not change the occupation number of $t_{2g}$ or $e_{2g}$ orbitals. Moreover, in the case of a parallel-spin $t_{2g}$-$e_g$ pair, the Coulomb interaction does not change the spin value of the orbitals. The exchange interaction between a $t_{2g}$ electron and a $e_g$ one of opposite spin is the only interaction which can flip $t_{2g}$ spins. If we neglect this interaction we can freeze the $t_{2g}$ electron variables and substitute them with an effective mean-field exchange interaction. This approximation should not alter too much the physics of the $MnO_2$ plane model which is dominated by $e_g$ electrons. The full many-body treatment can be extended to $t_{2g}$ by restoring the off-diagonal part of the $t_{2g}-e_{2g}$ exchange interaction at the expense of a higher computational cost, but this will not be done in this paper.

For oxygens we restrict to $p_y$ for the $(2i+1,2j)$ sites and to $p_x$ at the $(2i,2j+1)$ sites. These are the oxygen orbitals which play the major role in bridging the $Mn$ sites along the $x$ and $y$ directions. For symmetry reasons they are coupled to the $e_g$ sector only.

The Hamiltonian acting on the Hilbert space spanned by these orbitals is composed of the following parts

$$H = H_{bare} + H_{hop} + H_U^{Mn} + H_J^{Mn} + H_U^O \tag{24}$$

namely $H_{bare}$ which contains the one-particle energies of the orbitals, the hopping Hamiltonian $H_{hop}$ which moves electron between neighboring sites, the Hubbard correlations $H_U^{Mn}$, $H_J^{Mn}$ and $H_U^O$ for Manganese and Oxygen. The bare Hamiltonian, using letters $p$ and $d$ for the second quantisation operators on oxygen and manganese orbitals respectively, is:

$$H_{bare} = \sum_{i,j,g_d,s} (\varepsilon_d + (1/2 - s\, m_{i,j})h)\, d^\dagger_{g_d,s,2i,2j} d_{g_d,s,2i,2j} + \sum_{i,j,s} C_{api}\, d^\dagger_{3z^2-r^2,s,2i,2j} d_{3z^2-r^2,s,2i,2j} + (\varepsilon_p - U_p) \sum_{i,j,s} \left( p^\dagger_{x,s,2i+1,2j} p_{x,s,2i+1,2j} + p^\dagger_{y,s,2i,2j+1} p_{y,s,2i,2j+1} \right) \quad (25)$$

where the $m_{i,j}$ is a function of the site position and takes the values $\pm 1$ depending on the $Mn$ $t_{2g}$ magnetisation. The parameter $h$ is the exchange splitting induced on the $e_g$ orbitals by the $t_{2g}$ electrons. The index $g_d$ takes the values $g_d = x^2 - y^2, 3z^2 - r^2$. The $Mn$ one-particle energies $(\varepsilon_d + (1/2 - s\, m_{i,j})h)$ are spin ($s = \pm 1/2$) and position dependent and take into account the mean-field exchange with the $Mn$ $t_{2g}$ occupied orbitals $(xy, xz, yz)$( whose degrees of freedom are discarded from the model). The energy of the $3z^2 - r^2$ orbital is raised by a value $C_{api}$ above the bare energy to simulate the ligand-field splitting induced by the hopping to apical oxygens which are not otherwise treated in the model. The oxygen orbitals one-particle energy term takes into account the Hubbard coefficient $-U_p$ to compensate $H_U^O$ and favoring double and single occupations on oxygens.

The hopping term is

$$H_{hop} = t \sum_{i,j,g_d,s} \sum_{l=\pm 1} l \left( f_{g_d,x} p^\dagger_{x,s,2i-l,2j} d_{g_d,s,2i,2j} + f_{g_d,y} p^\dagger_{y,s,2i,2j-l} d_{g_d,s,2i,2j} + c.c. \right) \quad (26)$$

where

$$f_{3z^2-r^2,x} = f_{3z^2-r^2,y} = 1/2$$
$$-f_{x^2-y^2,y} = f_{x^2-y^2,x} = \sqrt{3}/2 \quad (27)$$

The Coulomb intra-site repulsive interaction for $Mn$ is given by $U_d$ for an electron pair on the same orbital, and by $U'_d$ for two different orbitals:

$$H_U^{Mn} = \sum_{i,j,g_d} U_d\, n_{g_d,s=+\frac{1}{2},2i,2j} n_{g_d,s=-\frac{1}{2},2i,2j} + \sum_{i,j,s_1,s_2} U'_d\, n_{3z^2-r^2,s_1,2i,2j} n_{x^2-y^2,s_2,2i,2j} \quad (28)$$

where letter $n$ denotes the number of electrons on a given orbital. The Coulomb exchange for $e_g$ orbitals is

$$H_J^{Mn} = J_d \sum_{i,j,s_1,s_2} d^\dagger_{3z^2-r^2,s_2,2i,2j} d^\dagger_{x^2-y^2,s_1,2i,2j} d_{3z^2-r^2,s_1,2i,2j} d_{x^2-y^2,s_2,2i,2j} \quad (29)$$

while the $e_g$-$t_{2g}$ exchange is included as a mean-field term inside $H_{bare}$.

Finally the oxygen Hubbard term is

$$H_U^O = \sum_{i,j} U_p \left( n_{p_x,s=+\frac{1}{2},2i+1,2j} n_{p_x,s=-\frac{1}{2},2i+1,2j} + n_{p_y,s=+\frac{1}{2},2i,2j+1} n_{p_y,s=-\frac{1}{2},2i,2j+1} \right) \quad (30)$$

To fix the free parameters of the model for the $MnO_2$ plane we use knowledge from our previous work on manganites[9]. Parameters are given in $eV$ units. The effective Slater integrals used in that work correspond, in the present model, to $U_d = 6.88$, $U'_d = 5.049$, $J_d = +0.917$. The exchange splitting $h$ induced by occupied polarised $t_{2g}$ orbitals is $h \simeq 2eV$. We use a hopping $t = 1.8$ taken from our previous work[9]. For oxygen we chose $U_p = 5eV$. The parameter $\varepsilon_p$ controls the amount of charge back-donation from oxygen to manganese. The predominant $O$ $2p$ character of doped holes found in manganites[10] corresponds to a value $\varepsilon_p$ which raises the bare oxygen orbitals energies above the bare $Mn$ ones. The value of $\varepsilon_p$ influences the average occupation of the $e_g$ orbitals. These occupancies match the ones found in the previous works for a value $\varepsilon_p \simeq 2$. We have fixed the $C_{api}$ effective apical ligand field splitting in the following way. We have considered an isolated plaquette formed by a $Mn$ site and the four neighbouring oxygen atoms. Using the present modelisation, with one $e_g$ electron, two electrons per oxygen, and the above parameters. For low values of $C_{api}$ the out-of-plane $e_g$ is occupied. We have increased $C_{api}$ till the in-plane orbital is occupied. This occurs at 1.5eV. Then we arbitrarily scale it down to $1eV$ in order to realise a system which, without the in-plane kinetic effect, which are enforced when the complete plane is considered, would have out-of-plane occupied $e_g$ orbital.

### 3.2. Parameters for the Hubbard model

In order to cross-check our implementation on the Hubbard-Model we compare it to the exact numerical calculation of a $4 \times 4$ system at half-filling with $U/t = 8$ (parameters of the Hubbard model). To obtain the Hubbard model from our $MnO_2$ plane model, we set to zero all the model parameters except $H_U^{Mn}$. Then we add an hopping term acting only between $x^2 - y^2$ orbitals of neighbouring sites :

$$H_{hop} = -t \sum_{i,j,s} \left( d^\dagger_{x^2-y^2,s,2i+2,2j} d_{x^2-y^2,s,2i,2j} + d^\dagger_{x^2-y^2,s,2i,2j+2} d_{z^2,s,2i,2j} + c.c. \right) \quad (31)$$

To span the same Hilbert as the Hubbard model we initialise the CC method with a ground state where the only occupied orbitals are the $d_{x^2-y^2}$ ones. The Hamiltonian, with the set of parameters here described, does not connect to other orbitals than the $d_{x^2-y^2}$'s and thus the effective degrees of freedoms coincide with those of the Hubbard model.

### 3.3. Parameters for the simple chain

We compare our method to exact solutions for a simplified system which we obtain in the following way. We consider spin-down electrons only, a one dimensional chain setting $N_y = 1$, and use the following parameters: $t = 1.8$, $C_{api} = 1$, $\varepsilon_p = 2$ and $\varepsilon_d, h, U_p, U_d, U_j = 0$. Calculation are done at different values of $U'_d$. The numbers of sites in the $x$ directions, $N_x$ is kept small enough so that an exact numerical solution, for the ground state and for the spectra, can be found.

## 4. DISCUSSION

### 4.1. The simple chain

We show in table (II) the reference state $|\Phi_0\rangle$ and some selected excitations for the simplest of our three studied models: the simple chain, for $U'_d = 3$. In the first column we mark the iteration number at which the excitation appears in the exponent $S_N$ (see equation (1)) when we calculate the ground state. The third column shows an Ascii-art view of the chain which can be read following the legend of table (I). The chain consists of six $MnO$ units and is populated with nine spin-down electrons. The occupied $Mn_{x^2-y^2}$ orbitals are represented by a cross composed of minus signs to signify a spin-down electron(see legend in table (I)) . In the reference state $|\Phi_0\rangle$, they are occupied at all six sites. The remaining three electrons populate one every two oxygen orbitals. These orbitals are represented, when they are occupied by a spin-down electron, by a minus sign. The spin-up sector is discarded in this simple model. The positions where excitations are created are highlighted by a black background. The ground-state energy, versus the number of excitations, is shown in figures (1,2 ), for two choices of the correlation energy: $U'_d = 1$ and $U'_d = 3$. We have let the algorithm to iterate long enough so that the per $Mn$-site energy converges to within few $meV$ from the exact energy. In figures (1,2 ) we plot the ground state energy versus the number of excitations, for $U'_d = 1$ and $U'_d = 3$. We reach a convergence error of the order of a hundredth of eV after 250 iterations for $U'_d = 3$ while for $U'_d = 1$ the convergence is faster as expected, due to the weaker correlation. In the one-particle case $U'_d = 0$ (not shown) the exact ground energy is recovered with 13 one-particle symmetrised excitations. We show in figure (3) the spectral function $A(q,\omega)$ (see equation (13)) for an ARPES operator acting on $Mn$ $3z^2 - r^2$ orbitals

$$C^{arpes}_k = \sum_{l<N_x} exp\left(i\frac{2\pi k}{N_x} l\right) d^\dagger_{3z^2-r^2,s=-\frac{1}{2},2l} \quad (32)$$

| | | |
|---|---|---|
| **a)** − − − − (x² − y², spin-down) | **b)** + + + | **c)** − + / − − + − / − + |
| **d)** × ××× × | **e)** + ++++ + × | **f)** ++[−]+ (black background) |

TABLE I. Guide to Ascii art representation of electronic states. **a)** a spin-down electron occupying a $x^2-y^2$ orbital. **b)** a spin-up electron in $z^2$ orbital. **c)** On the same $Mn$ site: a spin-down in $x^2-y^2$ and a spin-$x^2-y^2$up in $z^2$. **d)** the x symbol is used for double occupancy: here $x^2-y^2$ is doubly occupied. **e)** on the right of the $Mn$ site an oxygen orbital is doubly occupied. **f)** a black background highlights the orbital where an excitation is occurring. Here, with respect to a reference state given by e), a spin-down electron has been moved from the oxygen orbital to the $Mn$ $z^2$ orbital. Empty orbitals are not drawn.

| $N_{ecc}$ | $2^{nd}$ quantisation form | Ascii art |
|---|---|---|
| | $\vert\Phi_0\rangle$ | $\vert$ − − − − − − $\rangle$ |
| 1 | $d^\dagger_{3z^2-r^2_0} p_{x_1} \vert\Phi_0\rangle$ | $\vert$ [−] ■ − − − − − $\rangle$ |
| 2 | $p^\dagger_{x_{11}} d_{x^2-y^2_0} \vert\Phi_0\rangle$ | $\vert$ □ [−] − − − − − □ $\rangle$ |
| 3 | $d^\dagger_{3z^2-r^2_{10}} d_{x^2-y^2_8} \vert\Phi_0\rangle$ | $\vert$ − − − − [−] [−] $\rangle$ |
| 4 | $p^\dagger_{x_7} d^\dagger_{3z^2-r^2_6} p_{x_5} d_{x^2-y^2_6} \vert\Phi_0\rangle$ | $\vert$ − − − ■ [−] □ − − $\rangle$ |
| … | … | … |
| 300 | $d^\dagger_{3z^2-r^2_6} d^\dagger_{3z^2-r^2_4} p^\dagger_{x_3} d^\dagger_{3z^2-r^2_2} d_{x^2-y^2_2} d_{x^2-y^2_4} d_{x^2-y^2_6} d_{x^2-y^2_{10}} \vert\Phi_0\rangle$ | $\vert$ − [−] □ [−] [−] − [−] $\rangle$ |

TABLE II. Second quantisation and Ascii-art representation of some selected excitations, appearing in the $S$ operator which gives the ground state of the simple chain model, for $N_x = 6$ and $U'_d = 3$. The chain is populated with spin-down electrons The positions where second quantisation operators act on the reference state are marked with a black background. The $Mn$ sites are indexed with even indexes from 0 to 10, the $O$ sites with odd indexes from −1 to 11, with −1 being equivalent to 11 given the periodic boundary conditions.

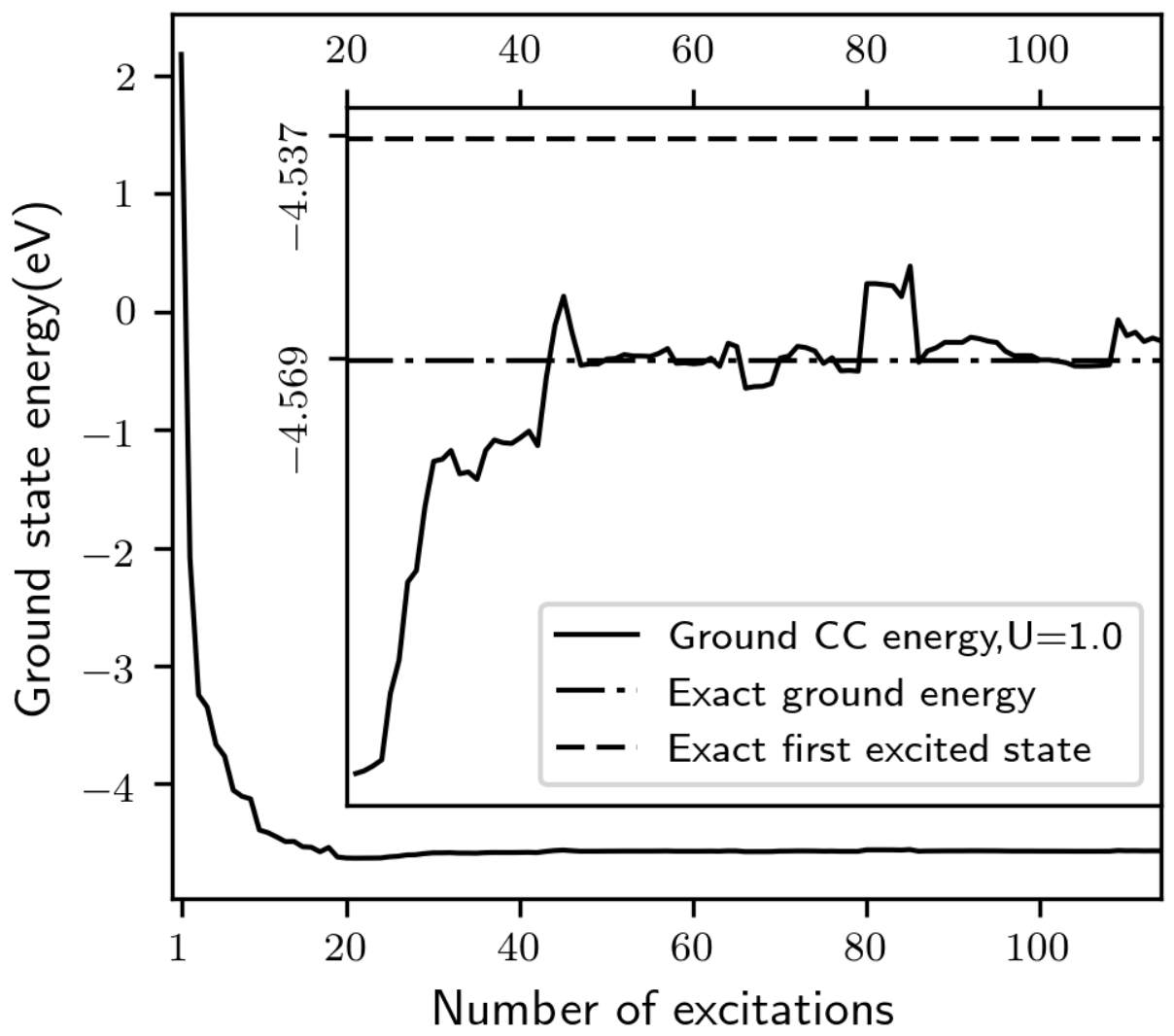

FIG. 1. Simple chain ground state energy as a function of the number of excitation in $S$ operator for $U'_d = 1$.

The SCCM spectral response , calculated with $1.5 \times 10^4$ spectral excitation operators, is shown for $U'_d = 1$, upper row, and $U'_d = 3$, lower row, and is compared to the exact calculation and to the Hartree-Fock mean field method. For $U'_d = 1$, at $k = 0$ the spectra consists in a unique peak at $\omega = 1.75$.

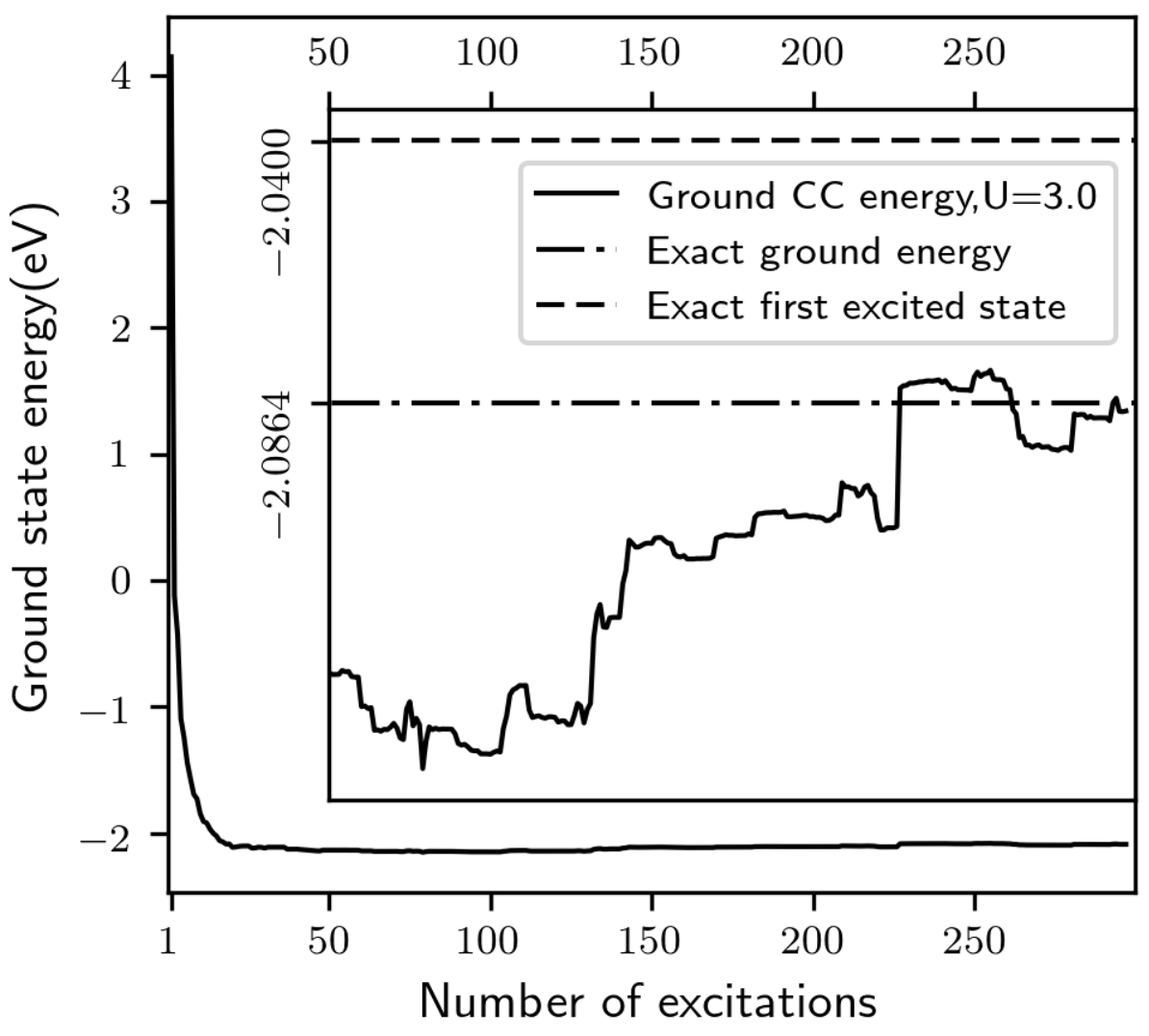

FIG. 2. Simple chain ground state energy as a function of the number of excitation in $S$ operator for $U_d' = 3$(right).

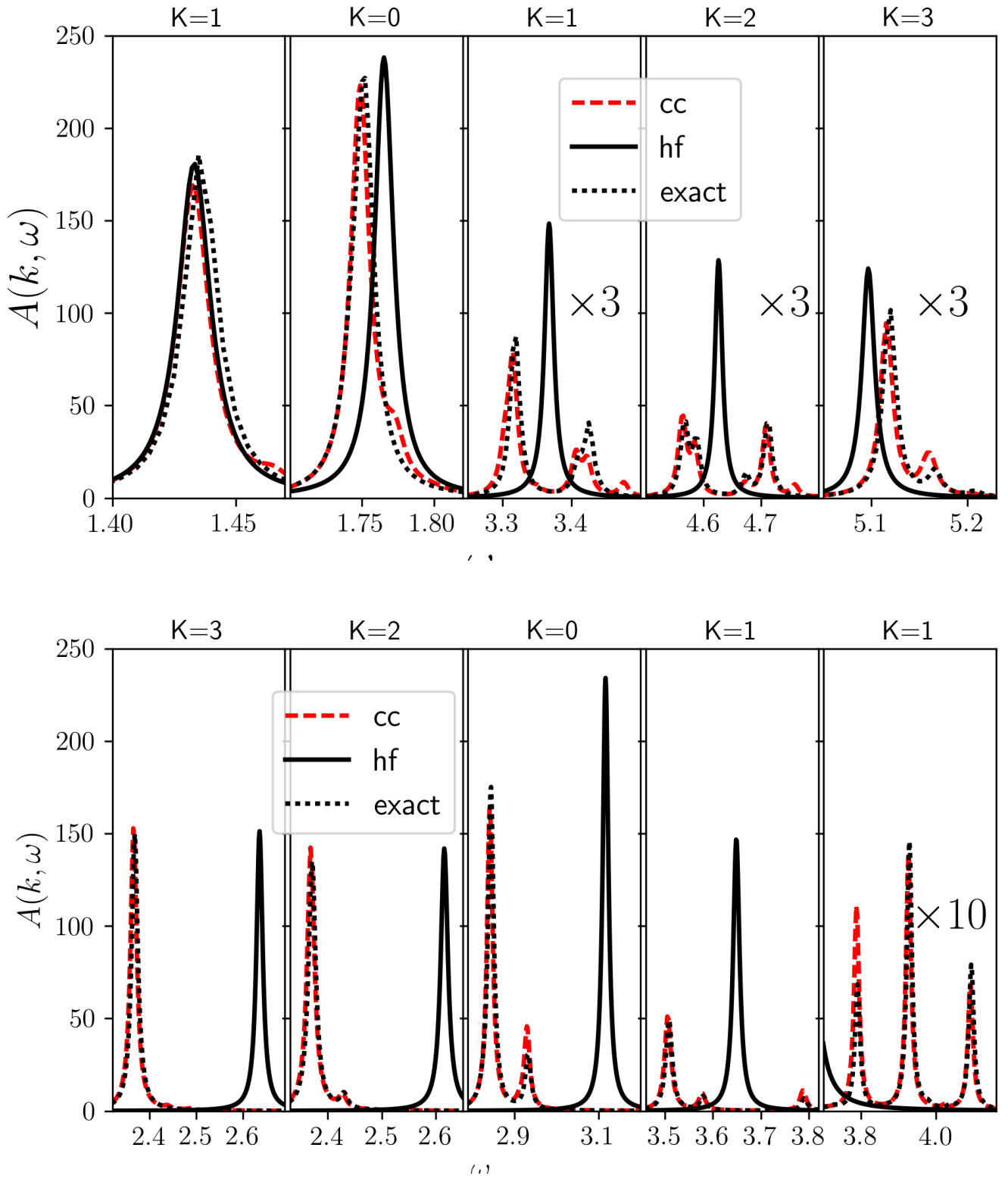

FIG. 3. The spectral function for the simple chain at $U_d' = 1$ and for $C_k^{arpes} = \sum_{l<N_x} exp\left(i\frac{2\pi k}{N_x} l\right) d^\dagger_{3z^2-r^2,\sigma=-\frac{1}{2},2l}$. Solid line: the mean field solution; dotted: the exact numerical solution; dashed(red color online): the SCCM solution.

At $k = 1$ this peak is found at lower energy and new spectral features appear around $\omega = 3.35$. At higher $k$ only these latter spectral features survive and shift to higher energy. Given the size of the chain, and time reversal symmetry, the spectra for $k = 4(5)$, not plotted, is identical to the one for $k = 2(1)$. For $U_d' = 3$ we show the main spectral features which account for more than 90% of the spectral density. The main spectral feature is found around $\omega = 2.9$ for $k = 0$ and moves to $\omega = 3.5$ for $k = 1$. Still for $k = 1$ we note the emergence of new spectral features at the slightly higher energy $\omega = 3.9$. At variance with the $U_d' = 1$ case these features keep moving at higher energy but, for $k = 2$ and $k = 3$, they remain marginally small and are not plotted. For these latter values of $k$ we note instead the persistence of the lowest energy peak which moves to lower energies. We can see from these numerical experiments that the SCCM reproduces well the shapes, the positions, and the strength of all the spectral features for all the considered values of $U_d'$, while the discrepancy of the Hartree-Fock method increases, as expected, increasing $U_d'$.

### 4.2. The Hubbard model at half filling and $U/t = 8$

.

#### 4.2.1. *Ground state with the CC method and by exact numerical solution.*

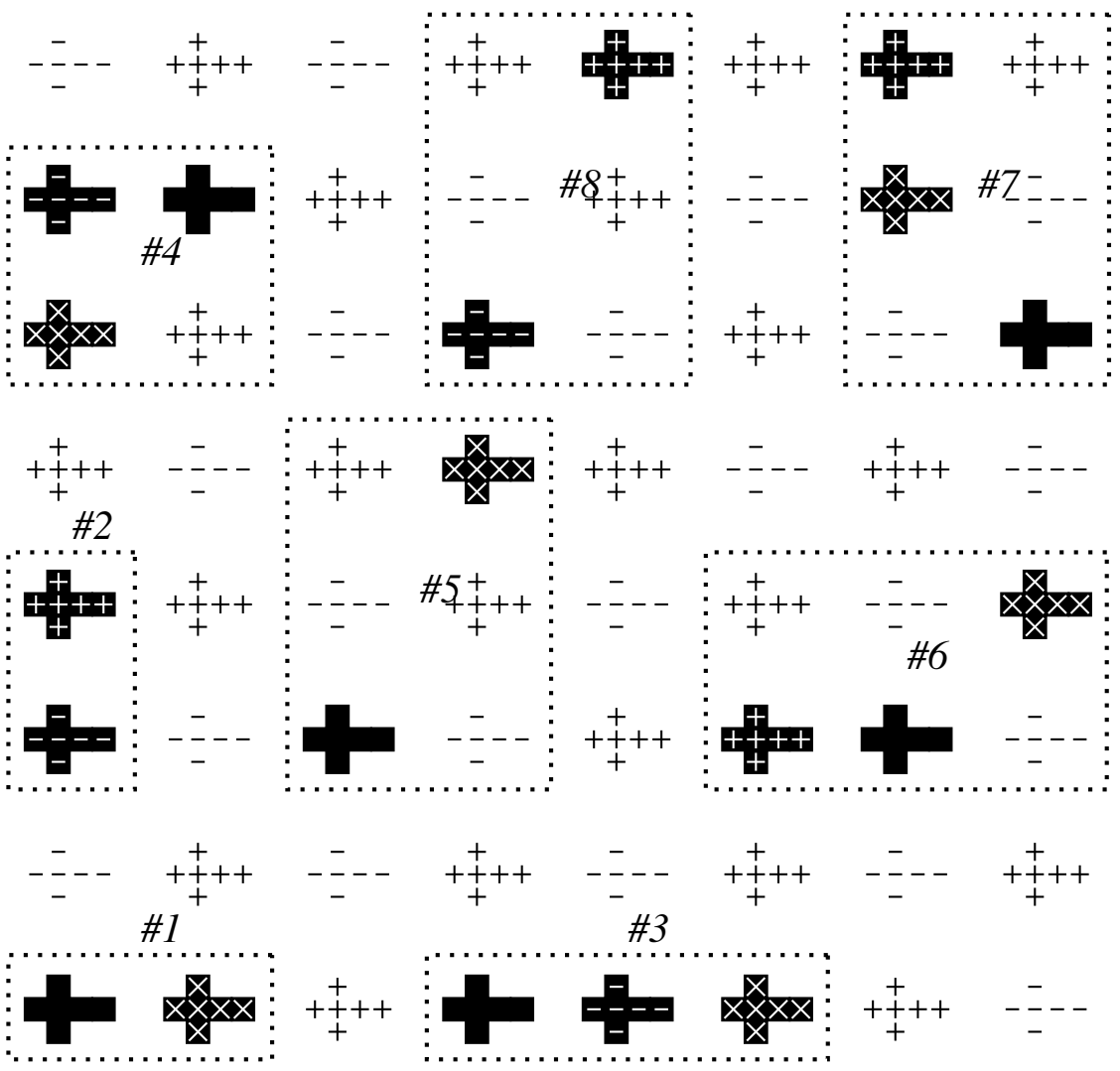

FIG. 4. Hubbard $8 \times 8$ model at half filling and $U/t = 8$. Ascii-art representation of the first eight excitations appearing in the $S$ operator for the ground state.

We show in figures (4) and (5), using ascii art, a selection of excitations appearing in the $S$ exponent for the ground state of the Hubbard $8 \times 8$ model at half filling, $U/t = 8$ and periodic boundary conditions. At each iteration step we introduce a new excitation and add it to the operator $S$ together with all the excitations that can be obtained by applying all the system translation, rotation and spin reversal symmetries.

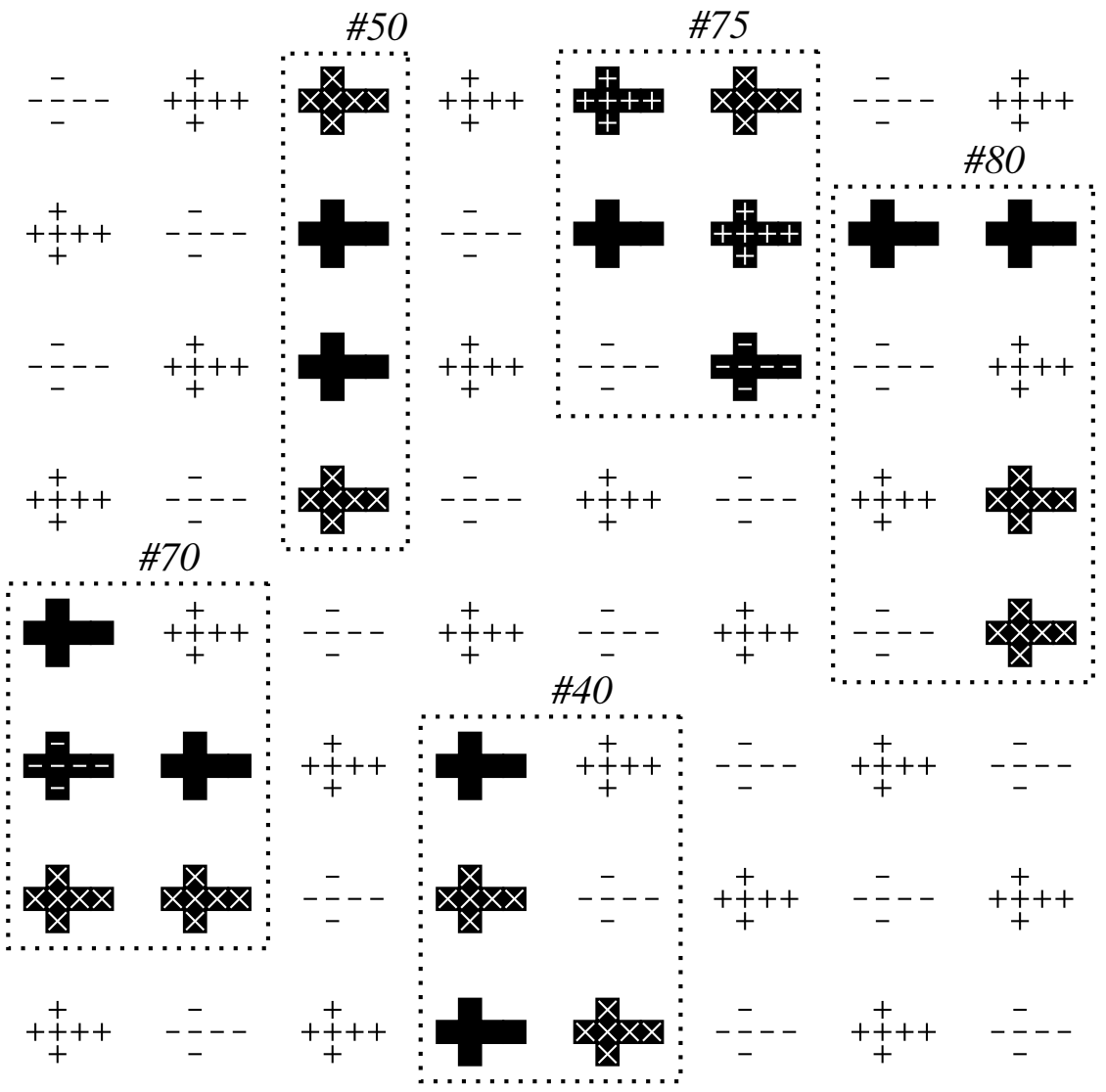

FIG. 5. Hubbard $8\times 8$ model at half filling and $U/t = 8$. A selection of excitations appearing at further stages of the convergence process.

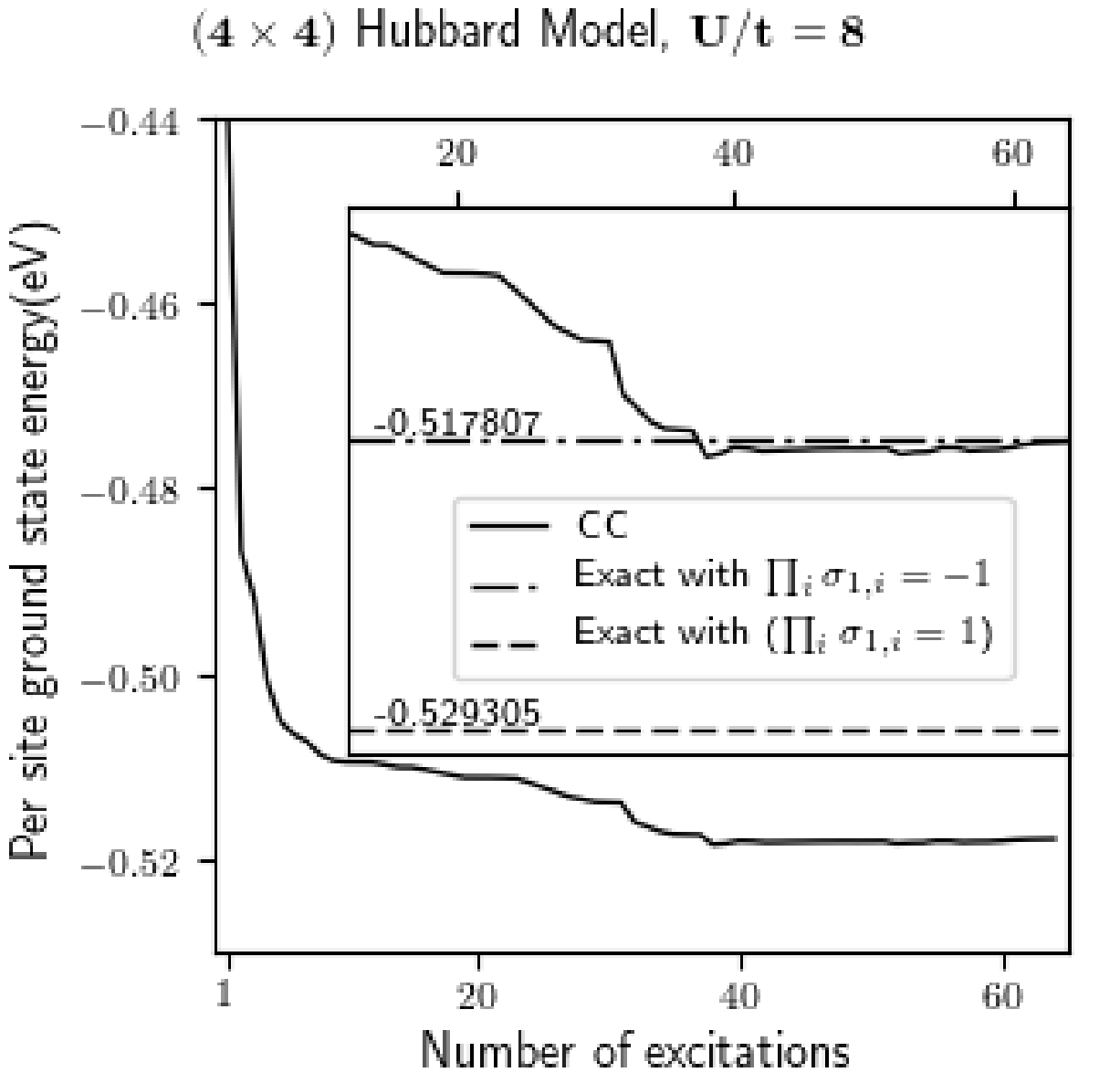

FIG. 7. Convergence for a smaller $4\times 4$ lattice and comparison with exact diagonalisation.

We show in figure (6) the ground state energy versus the number of excitations for the $8\times 8$ lattice and in figure (7) the comparison with exact diagonalisation for a $4\times 4$ lattice, whose smaller size allows for an exact numerical solution. The dimensionality of the Hubbard $4\times 4$ at half filling is $\sim 1.7\times 10^8$ which is still affordable with the current capability of a typical workstation. The numerical solution of the ground state is obtained by Lanczos diagonalisation where,for the first vector of the Krylov space, we take the same reference state that we have used in the CC method. If we restrain to the sector of even spin-reversal symmetry we still get the same ground state. The restriction is realised by starting the Krylov space with a vector which is the sum of the CC reference state plus the state obtained by swapping the spin-plus and spin-minus checker-boards. The Hamiltonian commutes with the spin-reversal operator and generates the spanned space which is therefore an eigen-space of the spin-reversal operator. Another ground state is found, instead, if we restrain to the odd spin-reversal symmetry sector, by starting the Krylov space with the difference between the two checker-board-swapped states. The non negligible energy difference between the two ground states indicates that collective spinonic fluctuations play an important role in the Hubbard model. The importance of these fluctuations is also visible in the asymptotic behaviour of the CC ground states energy. After reaching the energy region which corresponds to the collective spinonic excitation of the ground state, any further improvement of the CC ground energy is very slow. This is due to the fact that the collective spinonic fluctuations, appearing at all the possible wavelengths, require a large number of excitation operators, for all the sub-parts of the lattice concerned by a spin-reversal fluctuation, and each excitation concerns all the electrons inside these regions. We see instead, in figure (5), that the excitations appearing in $S$ are localised to small sub-parts of the lattice and thus cannot represent collective spinonic fluctuations. This happens despite the elevated number of iterations that we have considered and, in particular, in the region between the $40^{th}$ and $80^{th}$ term where the CC energy seems to be stabilised.

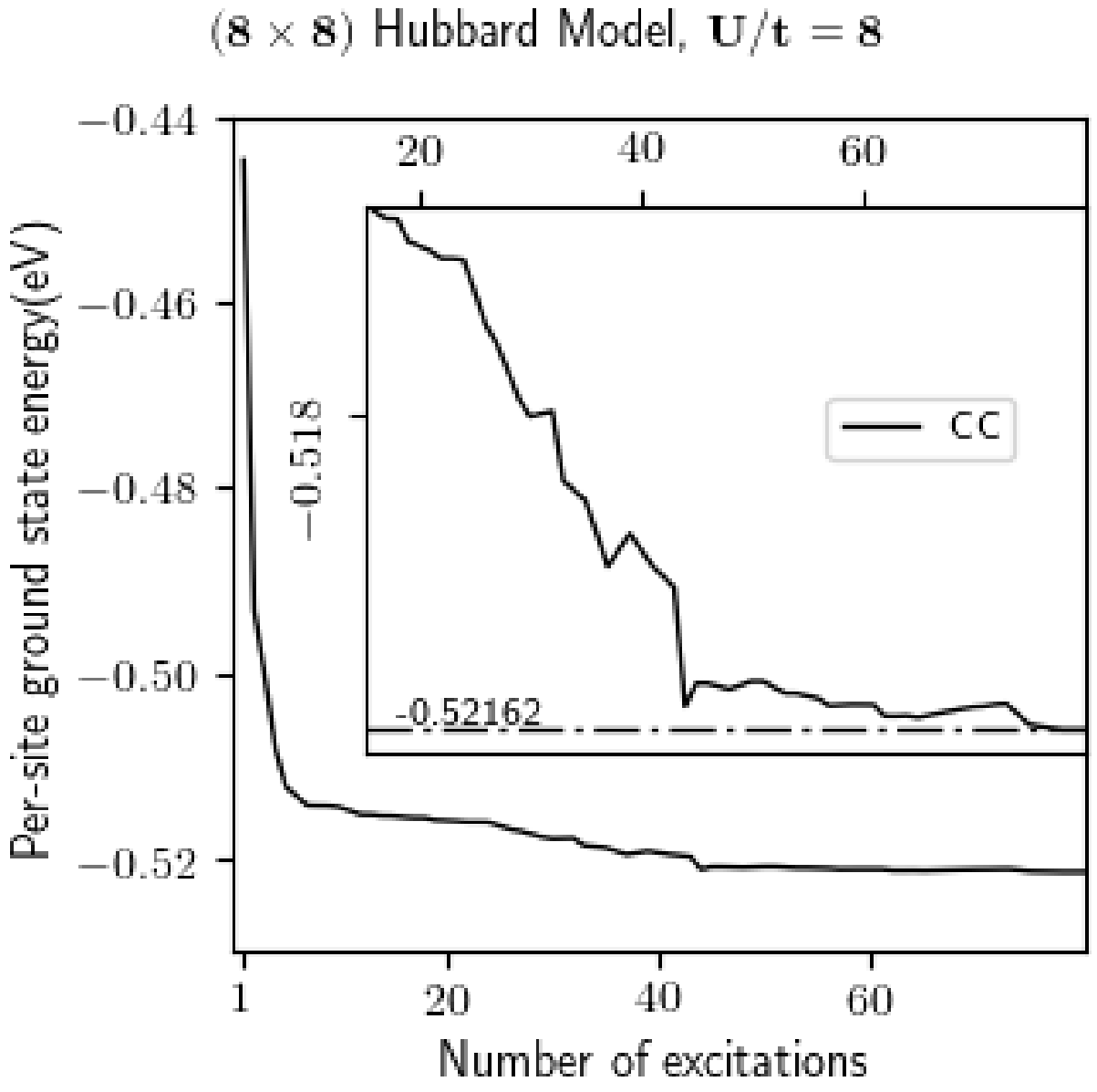

FIG. 6. ground state energy versus the number of excitations for the $8\times 8$ lattice.

We show in figure (8) the comparison between the spectra calculated in the framework of the exact numerical solution

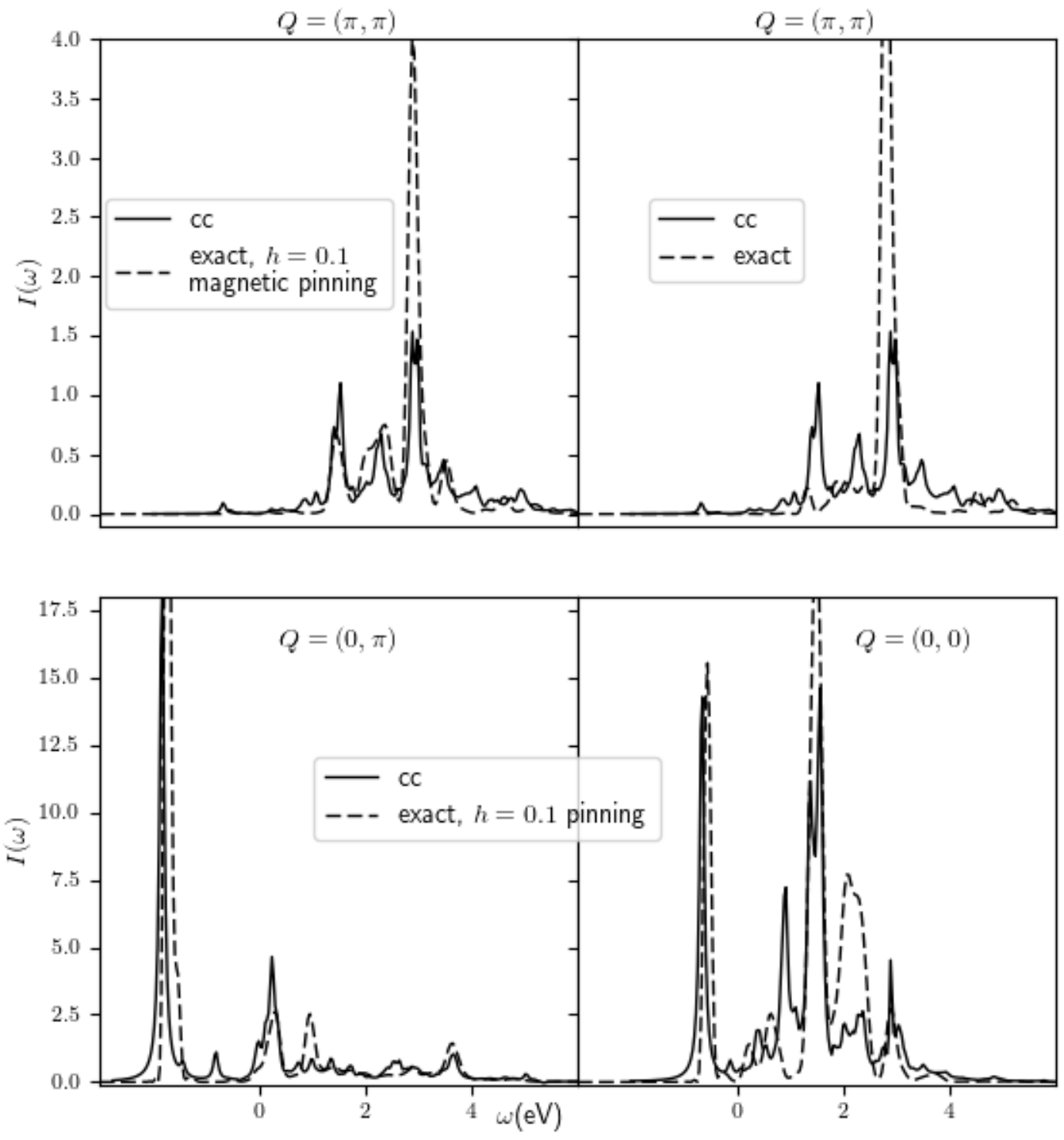

FIG. 8. The spectral function for the $4 \times 4$ Hubbard model at $U = 8, t = 1$ and for $C_k^{arpes} = \sum_{i_x,i_y} exp\left(i(Q_x i_x + Q_y i_y)\right) c_{2i_x,2i_y,\sigma=-\frac{1}{2}}$., calculated with $N = 16$ in $S_N$ and $3 \times 10^4$ terms in the resolvent $R_{\omega,\gamma}$.

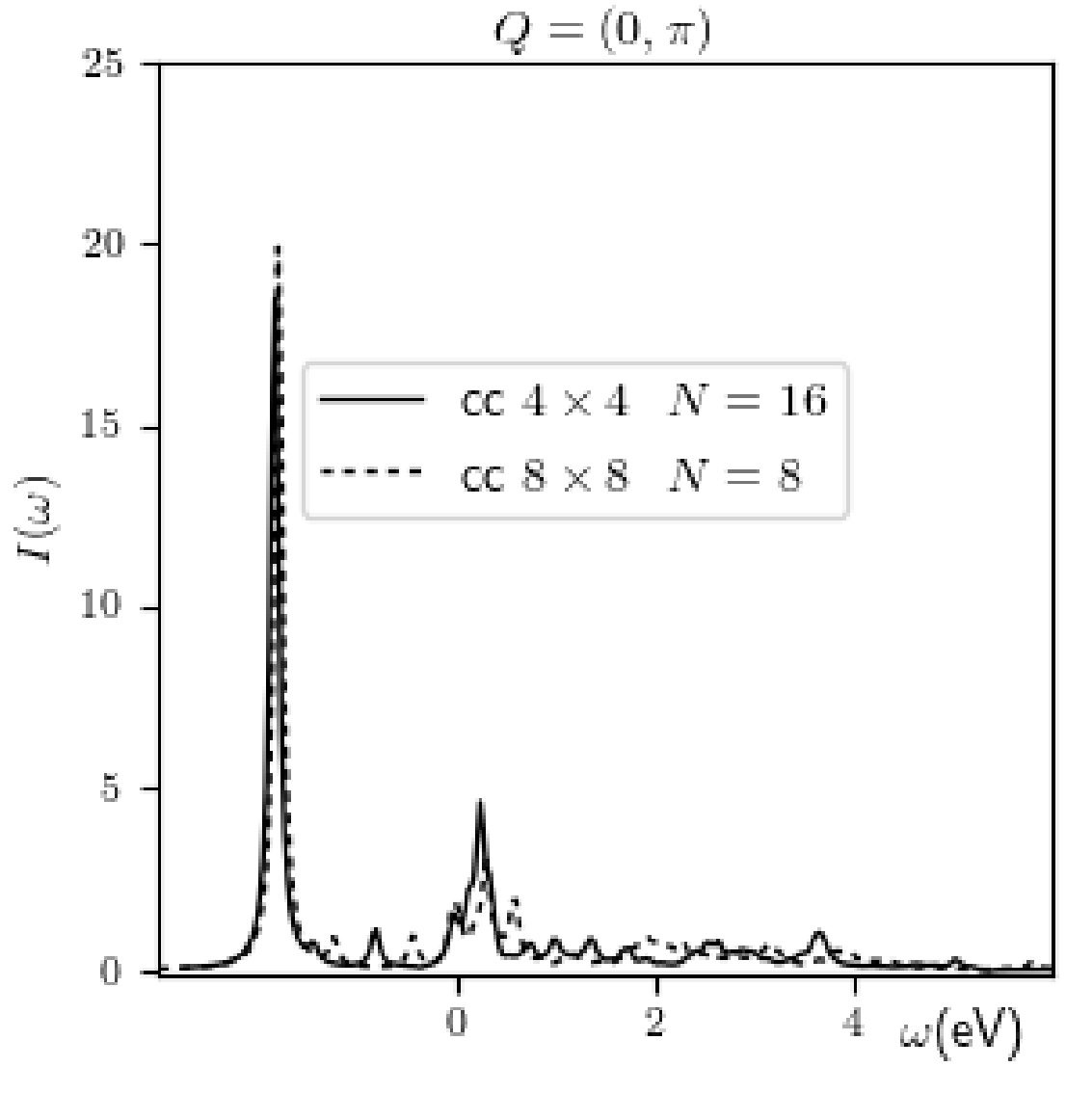

FIG. 9. The spectral function for $C_k^{arpes} = \sum_{i_x,i_y} exp\left(i(Q_x i_x + Q_y i_y)\right) c_{2i_x,2i_y,\sigma=-\frac{1}{2}}$, for the $8 \times 8$ Hubbard model with $U = 8, t = 1$. The SCCM calculation with $N = 8$ is compared to the SCCM calculation for the $4 \times 4$ model with $N = 16$. Both spectra have been calculated with $3 \times 10^4$ terms in the resolvent $R_{\omega,\gamma}$

and the spectra calculated by SCCM, for a $4 \times 4$ lattice. The exact numerical spectra was calculated applying the ARPES operator $C_k^{arpes} = \sum_{i_x,i_y} exp\left(i(Q_x i_x + Q_y i_y)\right) c_{2i_x,2i_y,s=-\frac{1}{2}}$ on the ground state obtained by Lanczos diagonalisation, and then obtaining the Chebychev polynomial coefficients of the spectra by iteratively applying the Hamiltonian on the resulting state[11]. In the Hubbard model there is just one type of active orbital, that in our implementation is $d_{x^2-y^2}$, and we represent it, in this context, with the symbol $c$ for notational simplification. The SCCM spectra has been obtained with 16 excitations ( $N = 16$ in formula (2)) and $3 \times 10^4$ terms in the resolvent ( formula (20)). We show in the upper row the spectra for $Q = (\pi, \pi)$. On the right graph, upper row, the SCCM spectra is compared to the exact one. We can see that apart for the main line position, many features in the SCCM spectra are in disagreement with the exact calculation. On the left side, the same SCCM spectra is compared to the exact calculation for a modified Hubbard model where an extra term $H_{pin}$ has been added to the Hamiltonian, in order to pin collective spinonic fluctuation. This term consists in a weak magnetic field acting on half of the sites: the checker-board black sites defined by $(i+j) \mod 2 = 0$, to favor spin-up occupation :

$$H_{pin} = \sum_{(i+j) \mod 2=0} -h\, c^\dagger_{2i,2j,1/2} c_{2i,2j,1/2} \tag{33}$$

and we have given this term the strength of $h = 0.1$ which is small compared to the hopping and correlation terms but can become non negligible for long-range fluctuations which swap large number of spin-up and spin-down sites. The effect of introducing this term is that, as expected, the modified system can be represented with improved fidelity by SCCM which starts, in our implementation, from a defined reference state with no fluctuations. The SCCM calculation at ARPES momentum $Q = (0, \pi)$ and $Q = (0, 0)$, shown on the lower line, reproduce the pinned Hubbard model with a similar level of fidelity. A larger number $N$ of excitations in the CC exponent $S_N$ could improve the fidelity of the SCCM calculation. However, despite the extensive properties of CC method which render it capable of treating larger system than the exact diagonalisation, for this specific example which is still treatable by exact diagonalisation, the CC requires important computational resources. Each excitation in the $S_N$ exponent is symmetrised by the system symmetry group, which for the half-filled $4 \times 4$ model consist of 128 elements, so the operator $S_N$ is the sum of thousands of terms already with $N$ as small as 16. The SCCM proceeds by keeping track of the residues in equation (21) which are generated by $e^{-S}(H - \omega - i\gamma)R_{\omega,\gamma}e^{S}$ where the resolvent $R_{\omega,\gamma}$ consists in our case of $3 \times 10^4$ terms. All this, traduced into our implementation, almost saturates the 250Gb memory of one node of the ESRF cluster. The situation can be alleviated by distributing the memory over several nodes. By an appropriate distribution and messaging scheme this profits also the execution time. However even with 4 nodes, which amounts to 112 cpu-cores, the execution time remain of the order of 24 hours. For comparison the exact numerical computation of the spectra with Lanczos diagonalisation and spectra decomposition in Chebychev polynomia completes on 28 cores within a walltime of 2 hours.

Considering the application to other systems, all this is nonetheless encouraging for the following reasons. First of all the size of the system we have considered is already very close to the limit of what can be done with the exact diagonalisation, but not of what we can do with the SCCM. As an example a $6 \times 6$ Hubbard system has a dimensionality of $8.2 \times 10^{19}$ which exceeds by far the memory capabilities of the most powerful super-computer nowadays, but we can show in figure (9) the SCCM spectra for the $8 \times 8$ Hubbard model calculated with $N = 8$ in $S_N$ and $3 \times 10^4$ terms in the resolvent. The number of excitations has been limited to $N = 8$, for the ground state, to limit the memory usage, because the string which represent the residues in second-quantisation are now four times longer. Despite the poor level of approximation of the ground state, which could be improved using more resources on a super-computer, the spectra still reproduces with good agreement the two main principal feature of the spectra. Moreover, the Hubbard model maybe considered as an extreme case for the application of the CC method, because long range fluctuations bring the system very far from the reference state. In the next section we are going to study the energy-loss spectroscopy of orbital-ordering in manganites for which we expect that these long range fluctuation be less severe. In these systems, in fact, the physics is dominated by the mobility of the $e_{2g}$ charge in the mean exchange field of the $t_{2g}$ electrons. As the long range magnetic ordering is well defined in this case, we build the CC modeling starting from a reference state which reproduces such ordering.

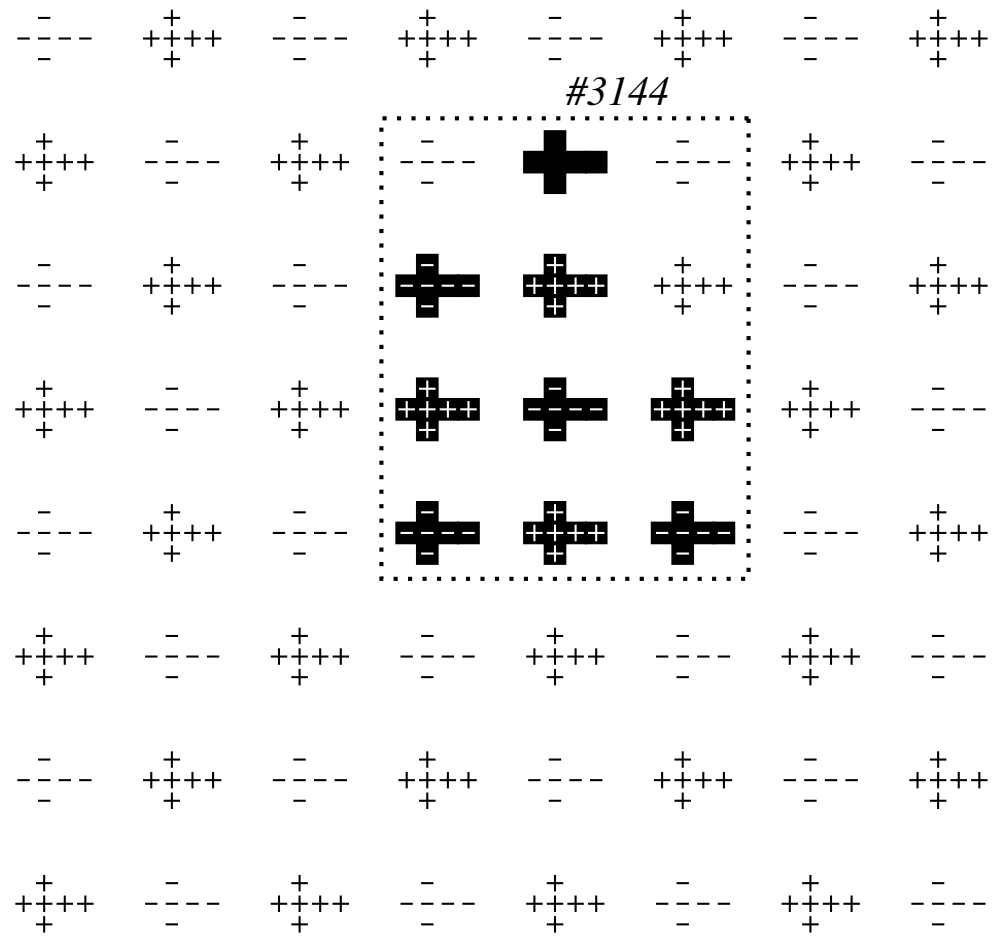

FIG. 10. An excitation occurring in the ARPES resolvent for the Hubbard system highlighting the role of collective spinonic fluctuation in dressing charge carriers.

#### 4.2.2. *Observation of a pairing mechanism*

The dimensionality of the Hubbard system is maximal at half-filling. By introducing one or more holes or electrons the dimensionality decreases. Our implementation of the Lanczos diagonalisation can cope with the half-filling case, it is *a-fortiori* suited to solve the problem where one or more holes, or electrons are introduced in the system. It is therefore interesting, as a side-study of our work on the Hubbard system, to calculate the ground state energy of the $4 \times 4$ Hubbard model by adding, to the half-filled system, one or two holes, with a determined total kinetic moment. As in the case of spin-reversal parity treated in the preceding sub-section, the total momentum of the system is determined by the first vector of the Krylov space. The ground energy of the half filled ground state is $E_0 = -8.468875$. Adding one hole we get the ground energy $E_{1h} = E_0 + \Delta_{1h}$ with $\Delta_{1h} = -1.678365$, while by adding two hole to the half-filled system we get $E_{2h} = E_0 + \Delta_{2h}$ with $\Delta_{2h} = -3.399965$. The ground state for the two-holes doped system is realised introducing a couple formed by a spin-up and a spin-down hole with $(\pi,\pi)$ total momentum. These energies satisfy the strict inequality $\Delta_{2h} < 2\Delta_{1h}$. We can now perform a conceptual experiment by considering an ensemble of $4 \times 4$ Hubbard systems at half-filling. By adding two holes, each in a different Hubbard system, we increase the energy by $2\Delta_{1h}$. The resulting total energy is therefore higher than what we obtain placing both holes on the same Hubbard system.

To characterise the fluctuations, given the calculated ground state of the $4 \times 4$ system, we calculate the expectation values of the following operator :

$$S_{AB} = \sum_{\alpha=x,y,z} S^{\dagger}_{A,\alpha} S_{B,\alpha}$$

with

$$S_{A,\alpha} = \sum_{(i_x+i_y) \mod 2=0} \sum_{s_1 s_2} \sigma^{\alpha}_{s_1 s_2} c^{\dagger}_{2i_x,2i_y,s_1} c_{2i_x,2i_y,s_2}$$
$$S_{B,\alpha} = \sum_{(i_x+i_y) \mod 2=1} \sum_{s_1 s_2} \sigma^{\alpha}_{s_1 s_2} c^{\dagger}_{2i_x,2i_y,s_1} c_{2i_x,2i_y,s_2} \quad (34)$$

where $\sigma$ are the Pauli matrices, and we are taking the spin-spin scalar product between the black and white sites of the checkerboard. For the nominal reference state, with spin-up black sites and spin-down whites sites, the expectation value is $< S_{AB} >= -64$. This is also the maximal value of the operator. The ground state of the half-filled system, calculated by exact diagonalisation is $< S_{AB} >= -60.4$. For the exact ground state of the one and two-holes systems we get $< S_{AB} >_{1h}= -42.5$ and $< S_{AB} >_{2h}= -30.8$ respectively. By naming $r$ the ratio between the $S_{AB}$ expectation values on the ground states and the largest magnitude values, which are 64, 56 and 49 for zero, one and two holes respectively, we get $r = 0.944$ , $r_{1h} = 0.76$ and $r_{2h} = 0.63$ for zero, one and two holes, while at the same time the total spin of the whole $4 \times 4$ lattice is found to be a singlet, a doublet, and again a singlet.

These numerical evidences highlight the following scenario:

- The collective spinonic fluctuations occurring in the ground state have a relatively long range so that, for the considered case, short range antiferromagnetic ordering, on the scale of the $4 \times 4$ lattice, is preserved($r = 0.944$).

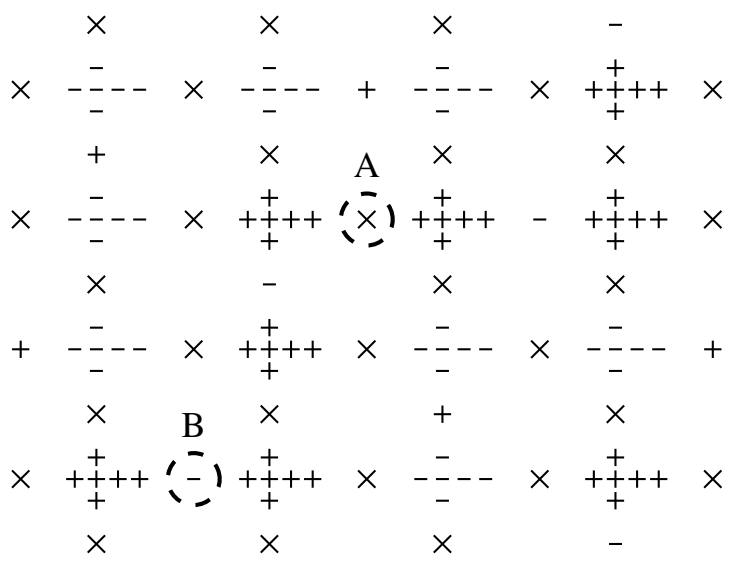

FIG. 11. Reference state for the $MnO_2$ plane at half-doping with orbital ordering. Along a ferromagnetic chain double-occupancy oxygen sites (called A sites) alternate with single occupancy sites (B sites). Two circles highlight one A site and a B site.

- Adding one hole to the half-filled system brings down $r$ to $r_{1h} = 0.76$. This is caused by the hole mobility which disturbs the antiferromagnetic ordering. The hole is dressed by a collective spinonic halo. Figure (10) shows one of the excitation concurring to the ARPES resolvent, for the Hubbard model, which highlights the role of collective spinonic fluctuation in dressing charge carriers.
- Adding one more hole brings down r to $r_{2h} = 0.63$, but the two holes move in a correlated way forming a singlet.

### 4.3. Modeling the half doped $MnO_2$ plane of $La_{0.5}Sr_{1.5}MnO_4$

Figure (11) shows, in ascii art (see legend in table (I)), the unit cell of the reference CC state used for the orbitally ordered ground state of a $MnO_2$ plane. The orbitals at $Mn$ sites are represented by symbols arranged in a cross pattern, for $x^2 - y^2$ orbitals, or a vertical line pattern for $3z^2 - r2$ ones (not present in the reference state). The orbitals at the oxygen sites are represented by a single symbol for the orbital which bridges between two $Mn$ sites. The symbols can be a minus sign, plus signs or the X symbol, for spin down, spin up, and double occupancy respectively (see legend in table (I)). This reference state corresponds to the orbital and magnetic ordering observed in the $MnO_2$ plane in $La_{0.5}Sr_{1.5}MnO_4$[9]. The half doping is realised, in this reference state, by removing one electron, every two oxygen sites, from the otherwise doubly occupied oxygen orbitals. In particular, we remove, in each chain, only electrons with majority spin. This is done in order to stabilise the ferromagnetic alignment along the zig-zag chains.

We show in figures (12,13), using ascii art, a selection of the excitations appearing in the S exponent for the ground state of the $8 \times 8$ $MnO_2$ model at half doping and periodic boundary conditions. At each iteration step we introduce a new excitation and add it to the operator S together with all the excitations that can be obtained by applying all translation, rotation and spin reversal symmetries of the reference state. These symmetry operations generators are: the translations generated by the $(2,2)$ and $(2,-2)$ vectors, the spin-reversal accompanied by a $(1,-1)$ translation, the swap of $x$ and $y$ axis accompanied by a $(1,1)$ translation. The energy, as a function of the number of resolvent terms, is shown in figure (14). We can identify three different regions. The fist one, for $N$ up to (about) 25, is a rapid descent of the energy and corresponds mainly to short range one-electron excitations. The second one, for $N$ between 25 and 90, corresponds mainly to short range two-particles and three-particles correlation terms, as can be checked in figure (13). The last part is the one where the energy has the lesser variations, and is increasingly contributed by many-body correlations of longer range, like excitations 134 and 198 (figure (13)) which correlates three $Mn$ and one $O$ atoms. The further we push the refinement process, the closer to the true ground state the CC solution is supposed to be. In particular the CC solution is supposed to restore the symmetry of the ground state, which is higher than for the reference state. A good indicator of this is the occu-pancy numbers at oxygen A and B sites (figure (11)). These are shown in figure (15) as a function of $N$. The A and B orbital occupations, which in the reference state are 1 and 2 respectively, should be equal to each other in the true ground state. This equivalence is realised, up to few percentages, beyond $N > 100$. The occupancy is evaluated at each $N$ by taking the derivative of the ground energy with respect to the $\varepsilon_p$ parameter.

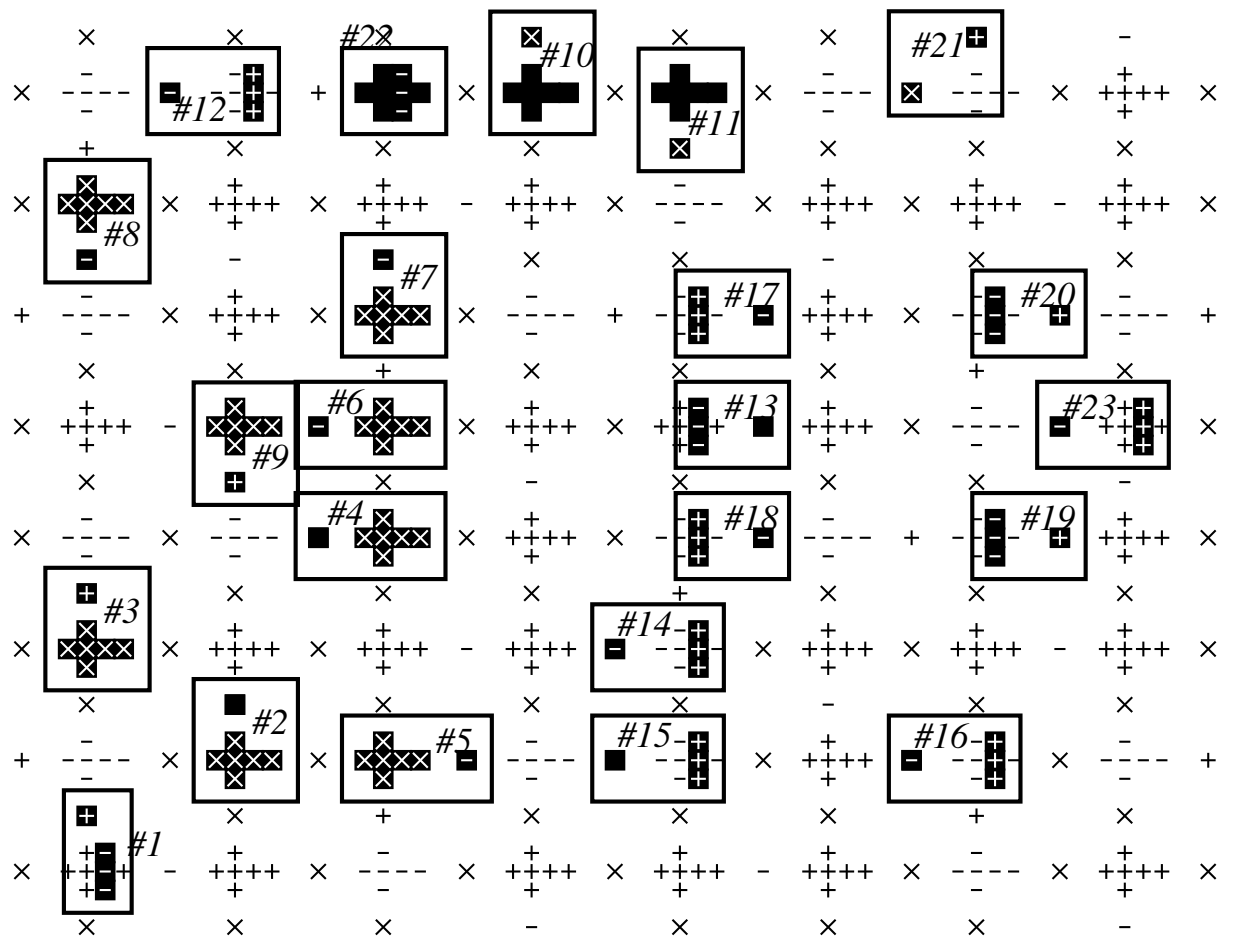

FIG. 12. First 23 excitations concurring to the $MnO_2$ plane ground state

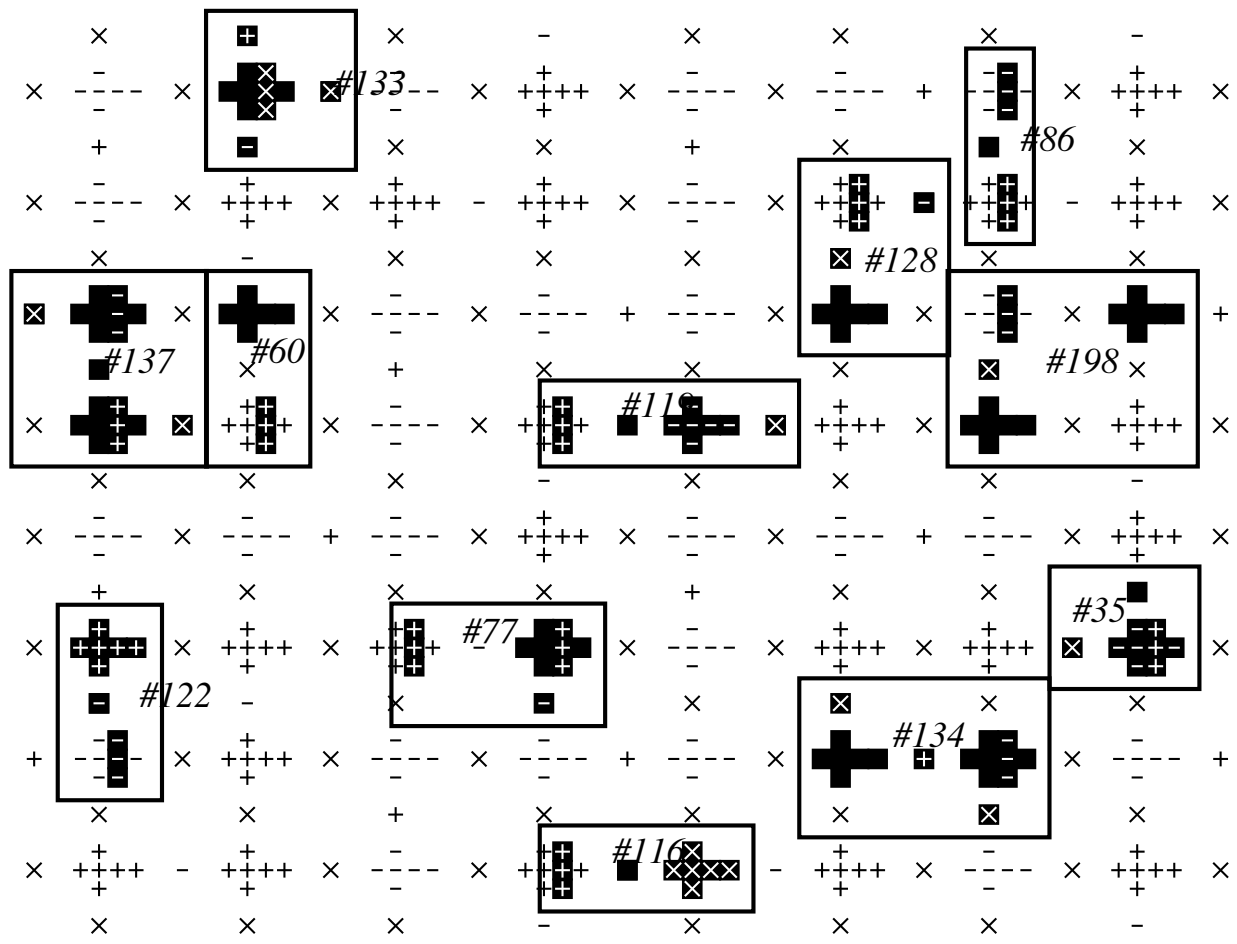

FIG. 13. A selection of excitations for further refinements of the $MnO_2$ ground-state.

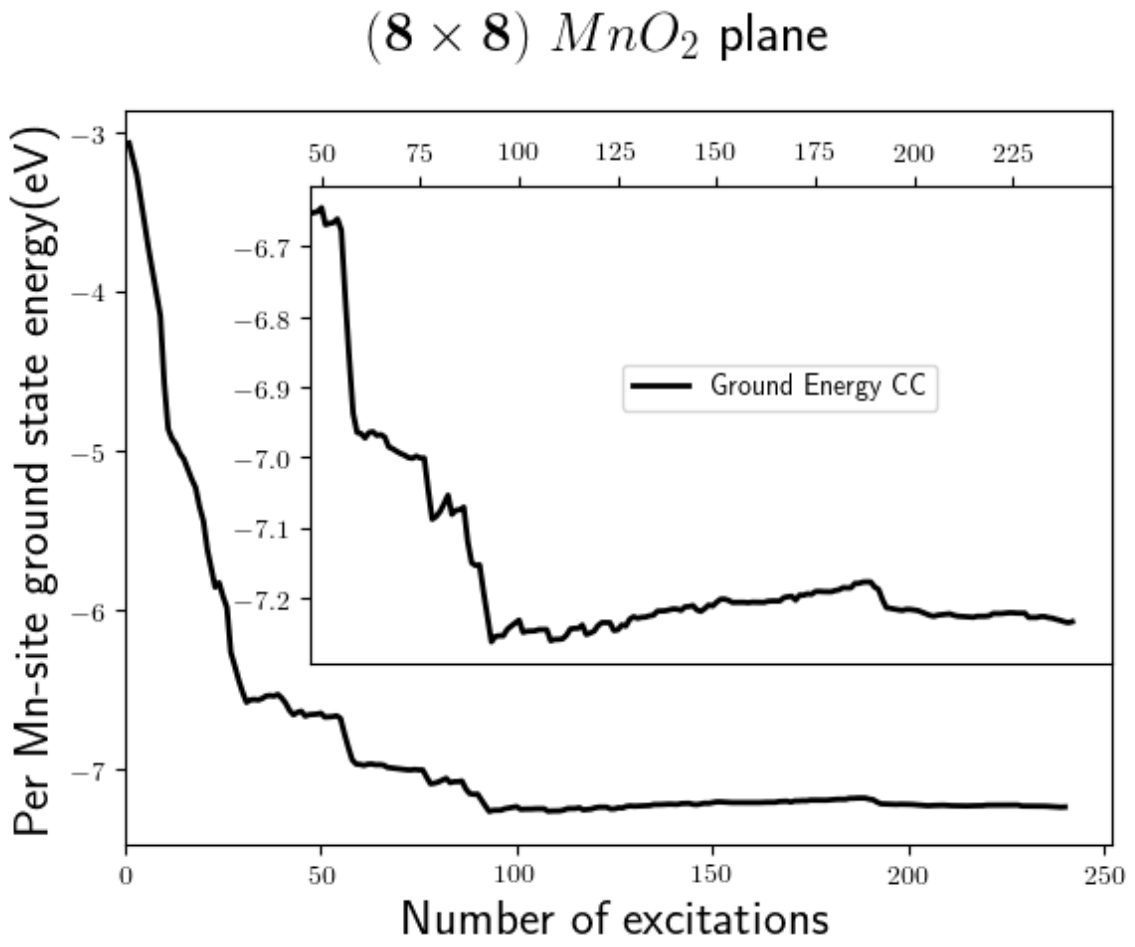

FIG. 14. Ground state per $Mn$-site energy of the $MnO_2$ plane as a function of the number of excitation in $S$ operator.

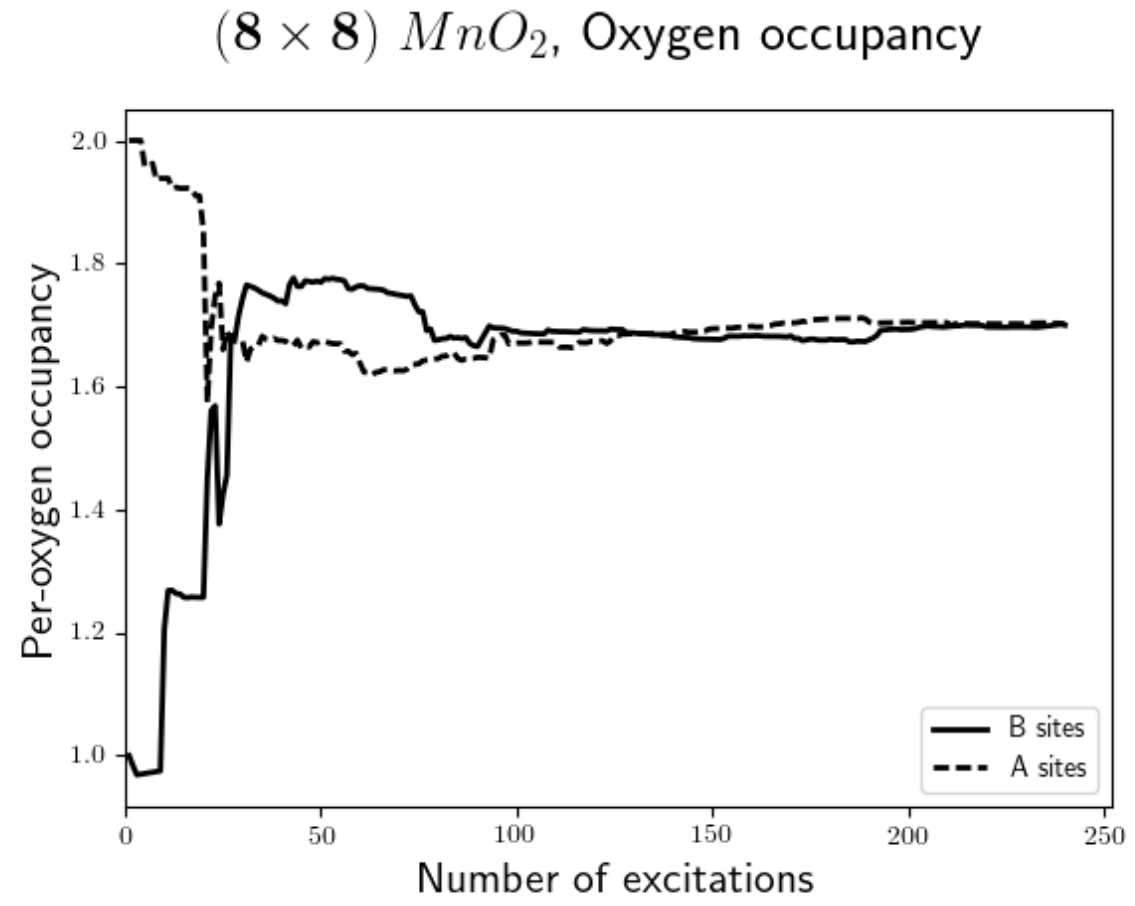

FIG. 15. Oxygen site occupancy for the $MnO_2$ plane as a function of the number of excitation in $S$ operator.

We have calculated the dynamic structure factor for the $MnO_2$ plane. To limit the computational cost we have calculated the $4\times4$ system at one $Q$ point: $Q = 2\pi/4$ which is the lowest non zero $Q$ available. We have kept only $N = 120$ excitations in the ground state. Our aim is to study eventual low energy excitations which have been observed in EELS[12] and have been attributed to oxygen-hole dynamic in the $MnO_2$ plane. We have calculated the resolvent for the $N_q^{\uparrow}$ operator only. The resolvent for $N_q^{\downarrow}$ can be obtained by symmetry.

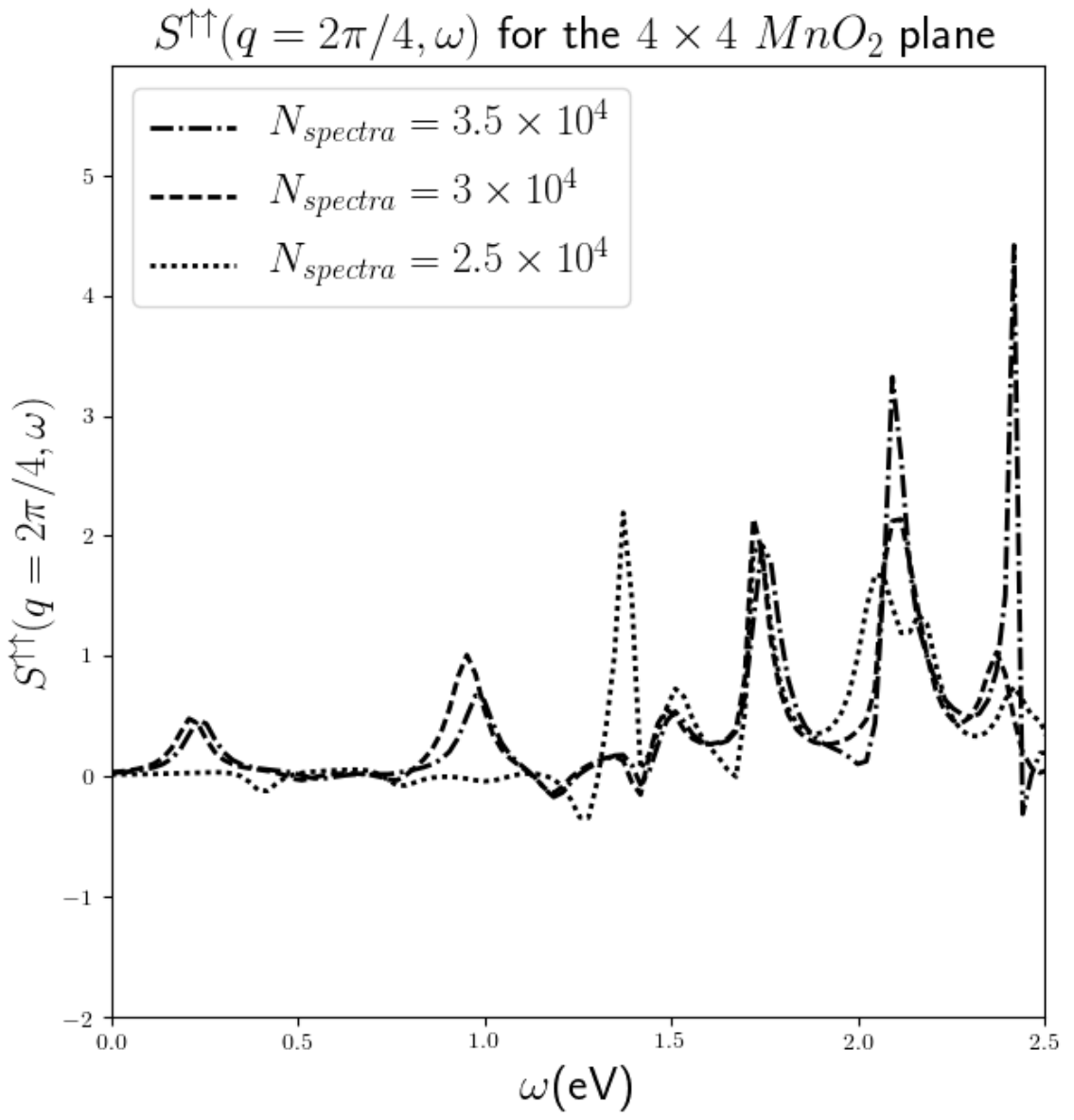

FIG. 16. The dynamic structure factor for the $MnO_2$ plane at different stages of the refinement process. At sufficiently high $N$ two low energy peaks seems to consolidate, one around 0.2eV, the other close to 0.9 eV.

The resulting spectra obtained, calculating $S^{\uparrow\uparrow}$ with SCCM, is shown in figure (16). We have pushed the refinement process far enough to see the emergence of two peaks at about 0.22eV and 0.95eV which seem to be stably reproduced for $N > 3.0\times10^4$. We have pushed our calculation as far as $N = 3.5\times10^4$.

In order to gain insight into the nature of these peaks we show in figure (17) the Green's function $G(B, N_q^{\uparrow}, \omega, q, \gamma)$ where $B$ is an arbitrary operator which we use to probe the nature of the spectral resonances. This Green's function can be obtained with negligible computational cost, once we have already calculated the dynamic structure factor from $G(N_q^{\uparrow}, N_q^{\uparrow}, \omega, q, \gamma)$ because both Green's function are given by the same resolvent(see equation (8)). We show with the dash-dotted line the modulus of $G(B^{\uparrow}, N_q^{\uparrow}, \omega, q, \gamma)$ with $B^{\uparrow} = p_{x,s,\uparrow,1,0}p^{\dagger}_{x,s,\uparrow,5,4}$. Such operator moves an oxygen hole (or electron) from one site to another of the same ferromagnetic chain. The dotted line is given instead by $B^{\uparrow} = d^{\dagger}_{2z^2-r^2,\uparrow,0,0}d_{x^2-y^2,\uparrow,0,0}$ and corresponds for an isolated $Mn$ atom to a crystal field excitation.

Our interpretation is the following. The lower peak at 0.22eV is a localised crystal field excitation and has little overlap with $P_q(p_{x,s,\uparrow,1,0}p^{\dagger}_{x,s,\uparrow,5,4})e^S|\Phi_0\rangle$. This explain why the resonance at 0.22eV is not visible with this probe. We interpret the resonance at 0.95eV as the promotion of an oxygen hole to an upper energy level. The counter-intuitive fact that the crystal probe $d^{\dagger}_{2z^2-r^2,\uparrow,0,0}d_{x^2-y^2,\uparrow,0,0}$ gives a more intense peak at the 0.95eV resonance than at 0.22eV is signature of a strong $Mn-O$ hybridisation and of a strong relaxation around the *Mn* site which hosts the crystal field excitation. This relaxation reduces the overlap of ligand orbitals between initial and final state for the crystal field resonance. We have also calculated the $B^{\downarrow} N_q^{\uparrow}$ for the crystal-field probe and we get again , in the low energy region, a peak at the same resonance position of 0.95eV and with a similar height. At the lower 0.22eV resonance position, instead, we still get a peak which is about half than for $B^{\uparrow} N_q^{\uparrow}$ but with negative sign. The negative sign is a further evidence that the 0.22eV resonance is a localised one. In fact in the limit of $q \to 0$, and at fixed $\omega$, the structure factor must go to zero. This because perturbing in the same way all electrons is equivalent to simply redefining the one-particle energies by a constant shift and this does not produces excitations. This means, because in our system $< BN_q >= 2 * (< B^{\uparrow} N_q^{\uparrow} + B^{\downarrow} N_q^{\uparrow} >)$ for spin-inversion symmetry, that the structure factor of the lower resonance is strongly suppressed, and this highlight the fact that the spatial shape of the resonance is smaller compared to the probe $q$ that we have used.

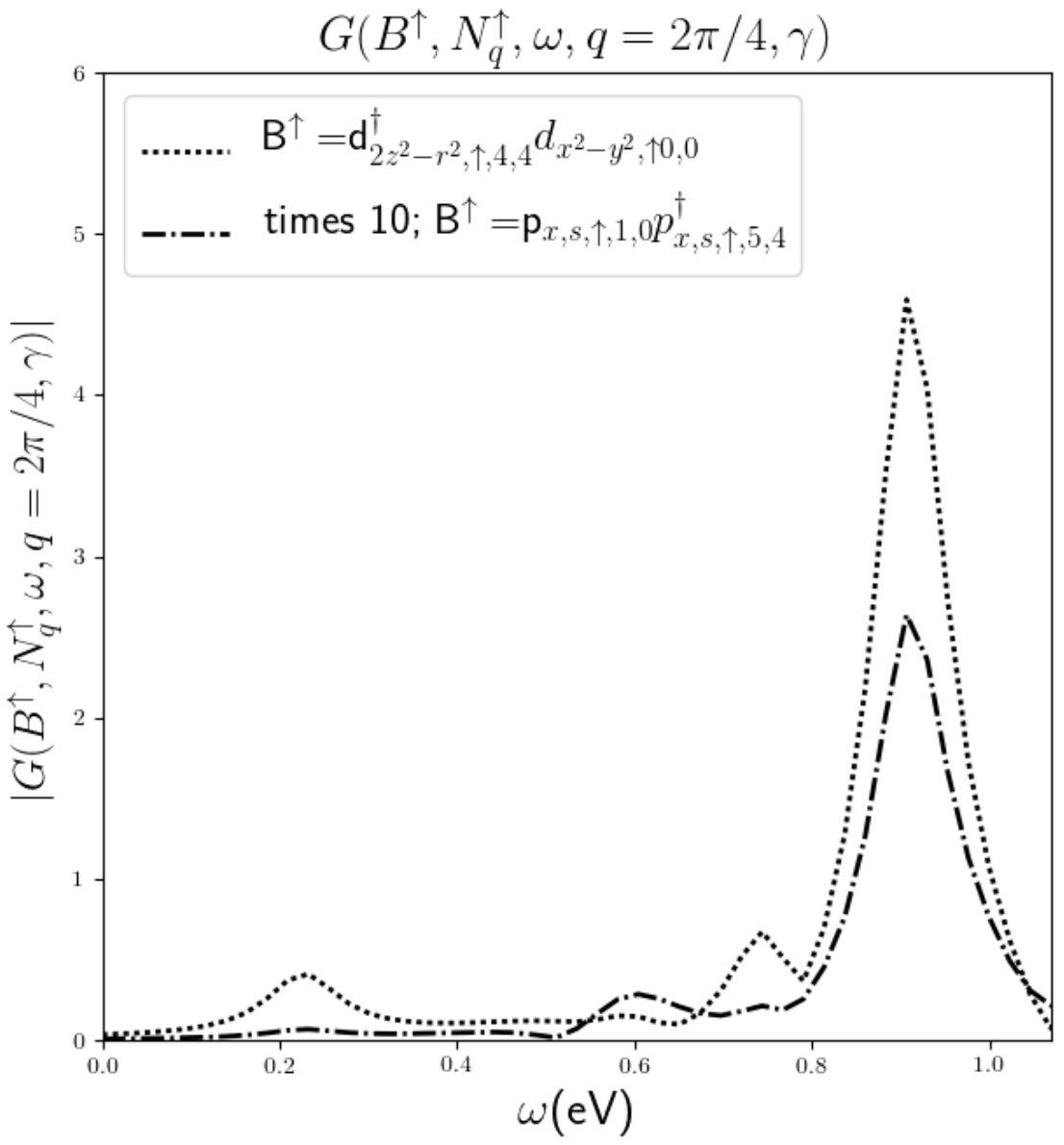

FIG. 17. The Green functions between the density operator and two ad-hoc probes : a probe acting on the *Mn* orbitals at a given site (dotted line) , a probe acting on the oxygen holes (dash-dotted line).

In an EELS experiment[12] on $La_{0.5}Sr_{1.5}MnO_4$, a low energy excitation around 1.1eV, observed also as a peak in the optical conductivity, for low $q$, has been observed, with a dispersive character, and has been attributed to the oxygen holes dynamics in the $MnO_2$ plane. We interpret our feature at 0.95eV as the simulation counter-part of this experimentally observed resonance.

As an outlook of further computational work that could be done to confirm our interpretation, the availability of more computing power would allow to perform calculations on a finer sampling of reciprocal space, using a $8 \times 8$ lattice, to check the dispersive character of the 0.95eV resonance, and the localised character of the 0.22eV resonance. Moreover an exploration of the parameter space could be done to improve the parameters choice and possibly reduce the energy position discrepancy between our simulation and the experiment. The above SCCM calculation represents a computational cost of 8 nodes, with 28 cores each, during 50 hours. The code is efficiently parallelised using a distributed memory scheme and can be run in best-effort mode, running during the unused time slots left empty by the scheduler of the computing cluster, each time restarting from the end of the previous step occurred before the last interruption. The calculation could be pushed further in the future according to the available computing power.

## 5. CONCLUSIONS

We have adapted the Spectral Coupled Cluster Method, previously derived from the Coupled Cluster linear response method for the calculation of spectra[5], to the calculation of generic Green's functions in the space-time Fourier transforms. We have numerically proved the validity of the formalism and its implementation, first, on a toy model and, then, on a $4 \times 4$ Hubbard model that, for comparison, we have diagonalised, and whose exact spectra we have numerically extracted. These preliminary tests have confirmed the validity of our method and, for the Hubbard model, have highlighted the role of the collective spinonic fluctuations on the ground energy, on the spectra, and a possible mechanism for pairing for which we have obtained a numerical evidence.

We have performed the calculation of the dynamic structure factor on the $MnO_2$ plane of $La_{0.5}Sr_{1.5}MnO_4$, and have observed two low energy resonances that, by numerical inspection, we have attributed to a crystal field excitation, the lowest energy one, and to a hole non-localised excitation the upper one. We have identified the latter with an experimentally already observed excitation[12].

The CC method is not plagued by the exponential growth of the computational complexity when the size of the system or the number of allowed excitations is increased. Therefore additional computing power can be fruitfully spent to gain further insight into the studied physical systems. We note that our method, which iteratively selects automatically the next CC excitation based on the values of the residue, produces, with higher priority, local excitations, so that further refinements can be focused more on increasing the excitation space, with lesser demand on the increase of the support lattice.

## 6. ACKNOWLEDGMENT

I thank Patrick Bruno and Markus Holzmann for critically reading the paper.